\documentclass[aps, pra, twocolumn,nofootinbib,superscriptaddress]{revtex4-2}
\usepackage[english]{babel}
\usepackage{amsmath,todonotes}
\usepackage{hyperref}
\hypersetup{colorlinks,allcolors=blue}
\usepackage{dsfont}
\usepackage{xcolor,ulem}
\usepackage{booktabs}

\newcommand{\beq}{\begin{equation}}
\newcommand{\eeq}{\end{equation}}
\renewcommand{\emph}{\textit}
\newcommand{\p}{{\rm p}}
\newcommand{\s}{{\rm s}}
\newcommand{\kk}{{\bf k}}
\newcommand{\KK}{{\bf K}}

\newcommand{\ii}{{\rm i}}

\definecolor{pinocolor}{HTML}{991111}

\definecolor{mycolorLorenzo}{HTML}{4ca916}

\usepackage{siunitx}
\usepackage{multirow}
\usepackage{placeins}

\begin{document}

\title{Optimal focusing conditions for bright spontaneous parametric down-conversion sources}
\author{Lorenzo Coccia}
\email{lorenzo.coccia@unipd.it}
\affiliation{Dipartimento di Ingegneria dell'Informazione, Universit\`a degli Studi di Padova, via Gradenigo 6B, 35131 Padova, Italy}
\author{Alberto Santamato}
\affiliation{CNIT, Via Moruzzi 1, Area di ricerca CNR, 56124 Pisa, Italy
}
\affiliation{Nu Quantum Ltd., Broers Building, 21 JJ Thomson Avenue, Cambridge, CB3 0FA, United Kingdom}
\author{Giuseppe Vallone} 
\affiliation{Dipartimento di Ingegneria dell'Informazione, Universit\`a degli Studi di Padova, via Gradenigo 6B, 35131 Padova, Italy}
\affiliation{Dipartimento di Fisica e Astronomia, Universit\`a degli Studi di Padova, via Marzolo 8, 35131 Padova, Italy}
\author{Paolo Villoresi}
\affiliation{Dipartimento di Ingegneria dell'Informazione, Universit\`a degli Studi di Padova, via Gradenigo 6B, 35131 Padova, Italy}
\affiliation{Padua Quantum Technologies Research Center, Universit\`a degli Studi di Padova, via Gradenigo 6B, Padova, Italy}
\affiliation{Istituto di Fotonica e Nanotecnologie–CNR, via Trasea 7, 35131 Padova, Italy}

\begin{abstract}
Optimizing the brightness of a spontaneous parametric down conversion (SPDC) source is an important task for many quantum information applications.
We investigate the optimal focusing conditions to maximize the number of photons produced in an SPDC process and coupled with single-mode fibers. 
We provide a general expression for the two-photon wavefunction, generalizing previous known results, by considering collinear and non-collinear emission. We present analytical expressions for our results in the thin crystal limit and clarify the relation between different focusing conditions already existing in the literature. Differently from what was previously reported, we show that the optimal ratio between
the pump waist and the generated photons waist depends on the emission angle: It is  $1/\sqrt2$ for collinear degenerate emission and approaches $1/2$ for larger collection angles. The role of spectral filters is also analyzed.
We support and enrich our discussion with numerical simulations, performed for type-I SPDC in a $\beta$ barium borate crystal. For this type of emission, we also investigate the role of the transverse walk-off outside the thin crystal regime.
\end{abstract}
\maketitle

\section{Introduction}
Spontaneous parametric down-conversion (SPDC) is one of the most commonly exploited physical mechanisms to generate quantum states of light and it has been used for fundamental tests of quantum mechanics, as well as for the realization of many protocols of quantum cryptography \cite{Scarani:2009zz,pirandola2020advances}, teleportation \cite{Bouwmeester1997}, and optical computing \cite{o2007optical,broome2013photonic}. The photons in this process are typically produced by the interaction of a laser pump with a non-linear optical medium that must have a rather strong second-order non-linear optical susceptibility $\chi^2$ to realize a bright source.
However, many other factors determine the properties and the number of emitted photons, such as  phase-matching and energy-conservation relations,  and the medium length and shape of the pump laser beam. The importance of these and other elements has been considered in many works with different types and degrees of approximation and we will not attempt to refer to them  exhaustively.  Instead, we are interested in a common experimental setup in which the photons involved in the SPDC process are emitted and collected by single-mode fibers. Mathematically, this configuration can be schematized by assuming spatial Gaussian modes for the photons, as has been done in several papers \cite{Bennink_note:_2010,Dragan2004,kolenderski_note:_2010,Ling2008,Andrews:04,BOVINO2003343,Boyd:68,Castelletto:05,Fedrizzi:07,Kurtsiefer,Ljunggren, Smirr2013}. In these works, different spectral properties of the emitted photons have been studied, using various perspectives or by invoking suitable approximations, such as the thin-crystal limit.

In this paper,  we will focus on the brightness of the emission process, evaluated as
\begin{equation} \label{eq:bright}
R_{\rm tot}= \int\left| \Psi(\omega_\ii,\omega_\s)\right| ^2 {\rm d} \omega_\ii {\rm d}\omega_\s \ ,
\end{equation}
where $\Psi(\omega_i,\omega_s)$ is the (unnormalized) wavefunction, whose explicit expression will be given in Eq.\ \eqref{eq:gen_psi}, which describes the frequency spectrum for the spatial Gaussian component of the SPDC photons, commonly denoted signal (s) and idler (i). 
The wavefunction $\Psi(\omega_i,\omega_s)$ is not normalized since it represents only the two-photon component of the full SPDC state (which includes also the vacuum and multi-pair components).
Experimentally, the integral in \eqref{eq:bright} gives the probability of detecting a pair of coincident photons when using single mode fibers to collect them. The choice of the domain of integration is related to the presence of filters, as we will discuss later.

It is well known that the brightness \eqref{eq:bright} depends on the waists of the photons in the process. According to the different assumptions made, however, different optimal values for the ratio between the pump waist and the signal or idler waists  have been reported in order to maximize \eqref{eq:bright} (see, e.g.,\! \cite{Bennink_note:_2010,kolenderski_note:_2010,Ling2008}). A clear relation between the different optimal values proposed is lacking.

Our article is part of this discussion, aiming to understand the conditions that lead to the optimal brightness in different regimes. Assuming Gaussian beams, we will tackle the problem from the momentum space perspective. We will first discuss how to make  a known paraxial approximation \cite{kolenderski_note:_2010} more accurate, providing an explicit formula, given in Eq.\ \eqref{eq:phi_fact_text}, to perform numerical simulations of the bi-photon wavefunction. Then, 
we will clarify the relation between values of the optimal ratio already found in the literature, relating them to different experimental situations: collinear or non-collinear emission, with or without spectral filters. Moreover, we will study the role of the transverse walk-off and find interesting optimal conditions.

To investigate different regimes and different geometrical configurations, we will perform numerical computations, choosing standard $\beta$ barium borate (BBO) crystals to realize type-I emission. A privileged role will be played by the SPDC sources based on crystals which are thin with respect to the focal depth of the pump field inside the crystal itself. In this particular configuration, we will derive analytical expressions to better understand the numerical results

The paper is organized as follows. In the next section we introduce the assumptions of our work. In Sec.\ \ref{sec:par_appr} we describe the paraxial approximation and derive formulas for the bi-photon wavefunction. In Sec.\ \ref{sec:thin} we discuss the thin-crystal limit and clarify the relation between the optimal focusing conditions proposed in the literature. In Sec.\ \ref{sec:numerical} we study corrections to the thin-crystal limit and report our numerical simulations. We summarize in Sec.\ \ref{sec:conclusion}. Some technical details are gathered in Appendixes \ref{sec:long_comp}-\ref{sec:computations}.

\section{Definitions and assumptions}
 The interaction Hamiltonian describing an SPDC process can be written as
\begin{equation}\label{eq:int_ham}
    \hat{\mathcal{H}}_I =\tilde{\zeta}^{(2)}_{\text{eff}} \int_V {\rm d}^3 \vec{r} \, \hat{D}_{\p}^{(+)}\hat{D}_{\s}^{(-)}\hat{D}_{\ii}^{(-)}+\text{H.c.} \ ,
\end{equation}
with $V$ the volume of the non-linear medium, which in the rest of the paper will be an uniaxial crystal. The quantity  $\tilde{\zeta}^{(2)}_{\text{eff}}$ is proportional to the inverse effective susceptibility and characterizes the strength of the interaction; in our discussion on the optimal focusing conditions it will be an irrelevant overall factor, but its exact value would instead be needed to obtain predictions about the absolute brightness. $\hat{ D}_a$, where $a=\p,\s,\ii$, are the displacement field operators of pump ($\p$), signal ($\s$), and idler ($\ii$) and their superscripts denote annihilation $(+)$ or creation $(-)$ operators. The three photons have a fixed but arbitrary polarization: If not stated otherwise, we will restrict the discussion to a specific type of SPDC (type I) only for numerical simulations. Note that we are writing the Hamiltonian in terms of $\hat{D}_a$ and not of the electric fields, since the latter choice would lead to some contradictions in the quantization procedure, as discussed in \cite{Quesada:17} (see also \cite{Schneeloch_2019} for  a review).

We consider a pulsed laser which propagates in the $z$ direction, impinging on a crystal of length $L$, centered at $z=0$, and oriented perpendicular to the $z$ axis. 
\begin{figure}
    \centering
    \includegraphics[scale=1.9]{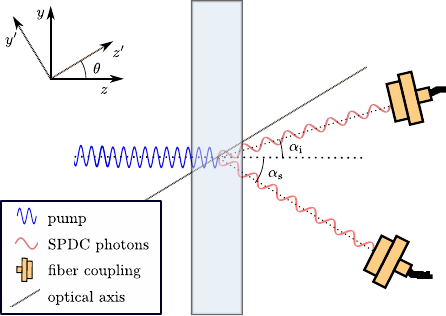}
    \caption{Experimental setup described in the text. The non-linear crystal of thickness $L$ is centered at $z=0$ . By $z'$ we denote the direction defined by the optical axis and by $y'$ its orthogonal direction in the $(y,z)$ plane.}
    \label{fig:setup}
\end{figure}
As shown in Fig.\ \ref{fig:setup}, we choose the $y$ axis so that the optical axis lies on the $(y,z)$ plane, forming an angle $\theta$ with the $z$ axis. The $x$ axis is defined so that
$(x,y,z)$ is a right-handed coordinate system. We parametrize the fields in the process selecting as independent variables the vacuum frequency $\omega$ and the wavevector components\footnote{Note that we use an arrow for the three-dimensional vectors and the bold font for the bi-dimensional vectors in the $(x,y)$ plane.} $\kk=(k_{x},k_{y})$. The explicit expression of the longitudinal component $k_z(\omega,\mathbf{k})$ can then be obtained using the dispersion relations, as discussed in Appendix \ref{sec:long_comp}. Parametrizations analogous to ours can be found in \cite{Rubin94, kolenderski_note:_2010, Karan_2020}. 

As is customary, we consider a strong pump field so that we can ignore depletion and we can write it as a classical field
\begin{equation}
    {D}_\p^{(+)}(\mathbf{r},t)\hspace{-0.1cm}=\hspace{-0.1cm} \frac{\mathcal{D}_\p}{(2 \pi)^{\frac{3}{2}}}\hspace{-0.1cm} \int {\rm d}^2 \mathbf{k}_{\p} {\rm d}\omega_\p \mathcal{A}_\p(\omega_\p,\kk_\p) e^{-i\vec{k}_\p\cdot \vec{r}}e^{-i\omega_\p t} \hspace{-0.1cm} .
\end{equation}
Here, $\vec{r}=(x,y,z)$, $\vec k_\p=(\mathbf{k}_\p,k_{\p z}(\omega_p,\mathbf{k}_\p))$, and
$\mathcal{D_\p}$ parameterizes the amplitude of the field. We also assume the factorable form between temporal and spatial components
\begin{equation}\label{eq:pump_fact}
\mathcal{A}_\p(\omega_\p,\kk_\p)=\mathcal{A}_\p^{\rm temp}(\omega_\p)
u_\p(\kk_\p)
\end{equation}
with a temporal normalized wavefunction given by $\mathcal{A}_\p^{\rm temp}(\omega_\p)$ and a spatial normalized wavefunction given by
$u_\p(\kk_\p)$. For the signal photon produced in the SPDC process we have
\begin{equation}
    \hat{{D}}^{(-)}_{\s}(\mathbf{r},t)=\frac{\mathcal{D}_{\s}}{(2 \pi)^{\frac{3}{2}}}\hspace{-0.1cm}\int {\rm d}^2 \mathbf{k}_{\s} {\rm d}\omega_{\s} \  e^{i\vec k_{\s}\cdot \vec{r}}e^{i\omega_\s t} \hat{a}^{\dagger}(\mathbf{k}_{\s},\omega_\s)
\end{equation}
and similarly for the idler, replacing $\s$ with $\ii$. 

Keeping only the first order in the expansion of the evolution operator, the (unnormalized) bi-photon wavefunction, describing the photons emitted in the process, is given by 
\begin{equation}\label{eq:general_wave}
    \left \lvert \Psi \right \rangle =-\frac{i}{\hbar}\int_{-\infty}^{\infty}{\rm d}t \  \hat{\mathcal{H}}_I(t) \left \lvert \text{initial} \right \rangle \ .
\end{equation}
Assuming that the photons are created in spatial modes described by $u_\s(\mathbf{k}_\s,\omega_\s)$ and $u_\ii(\mathbf{k}_\ii,\omega_\ii)$, we can easily find the expression (see also \cite{kolenderski_note:_2010})
\begin{equation}\label{eq:gen_psi}
\Psi(\omega_\ii,\omega_\s)=
\mathcal N
\mathcal{A}_\p^{\rm temp}(\omega_\ii+\omega_\s)\Phi(\omega_{\ii},\omega_{\s})
\end{equation}
where, rescaling the $z$ coordinate as $z= L Z/2$,
\begin{equation}
\label{PhiGeneral} 
\begin{split}
\Phi(\omega_{\ii},\omega_{\s})&=\frac{L}{2}
\int {\rm d}^2\kk_{\ii}{\rm d}^2\kk_{\s}
\int_{-1}^{1}\!\!\!\!\! {\rm d}Z\, u_{\p}(\kk_{\ii}+\kk_{\s})
\times
\\ &\times u^*_{\s}(\kk_{\s},\omega_{\s}) u^*_{\ii}(\kk_{\ii},\omega_{\ii})e^{-i\frac L2Z\Delta k_{z}(\kk_{\ii},\omega_{\ii};\kk_{\s},\omega_{\s})} \ .
\end{split}
\end{equation}
To derive this formula, we integrated over the time and the transverse spatial coordinates, obtaining the conditions $\omega_\p=\omega_\ii+\omega_\s$ and $\mathbf{k}_{\p}=\mathbf{k}_{\ii }+\mathbf{k}_{\s}$ in the form of delta functions. Note that we also consider the possibility of spatial poling, so that the normalization constant $\mathcal{N}$ in \eqref{eq:gen_psi} is written in terms of the amplitudes of the displacement fields as $\mathcal{N}= \tilde{\zeta}^{(2)}_{\text{eff}} \mathcal{D}_\p \mathcal{D}_\ii \mathcal{D}_\s G_m/ i \hbar$, with $G_m$ the Fourier coefficient of the spatial poling distribution of order $m$ \cite{Boyd}. 
Calling $\Lambda$ the spatial poling period, at order 
 $m$
the phase mismatch $\Delta k_{z}$ is
\begin{equation}\label{eq:deltakz}
\begin{split}
\Delta k_{z}(\kk_{\ii},\omega_{\ii};\kk_{\s},\omega_{\s})
=&m\frac{2 \pi}{\Lambda}+
k_{\p z}(\kk_{\ii}+\kk_{\s},\omega_\ii+\omega_\s)+
\\
&-k_{\ii z}(\kk_{\ii},\omega_\ii)
-k_{\s z}(\kk_{\s},\omega_\s) \ ,
\end{split}
\end{equation}
where $k_{a z}$ are the longitudinal components of the wave vectors inside the crystal. Unpoled crystals correspond to $m=0$.

For uniaxial crystals, the explicit form of $k_{az}(\kk_a,\omega_a)$ depends on the polarization of the corresponding wave. Writing the ordinary and the extraordinary refraction indices as $n_o$ and $n_e$, respectively, the refraction index at an angle $\theta$ to the optical axis is
\begin{equation}
    \frac{1}{n^2_\theta}=\frac{\sin^2\theta}{n_e^2}+
\frac{\cos^2\theta}{n_o^2} \ .
\end{equation}
We can write
\begin{equation}
\label{kz_general}
k_{z}(\kk,\omega)
=\beta k_y
+\sqrt{\left(\frac{n\omega}{c}\right)^2-(\gamma k_x)^2-\left(\gamma\frac{n}{n_o}k_{y}\right)^2} \ ,
\end{equation}
where 
$\beta=(\gamma^2-\frac{n^2}{n^2_o})\sin\theta\cos\theta$
and the values of the parameters depends on the polarization as follows:
\begin{align}
\label{gamma_ext} \quad
n=n_{\theta},\quad\,\,
\gamma=\frac{n_\theta}{n_{e}} \quad &{\rm (extraordinary)} \ ,
\\
\label{gamma_ord} \quad n=n_{o},\quad\,\,
\gamma=1 \quad &{\rm (ordinary)} .
\end{align}
More details can be found in Appendix \ref{sec:long_comp}. Note that the dependence on $\omega$ is also implicit in the indices of refraction $n_o(\omega)$ and $n_e(\omega)$
and thus in $\beta(\omega)$, $n(\omega)$, and $\gamma(\omega)$. According to the type of phase-matching considered,
it is necessary to express the longitudinal component of the wavevectors $k_{az}$ by using
Eqs.\ \eqref{kz_general} and  \eqref{gamma_ext} or \eqref{gamma_ord}.

The above-derived two-photon wave function \eqref{eq:gen_psi} is very general. We will now focus on some experimentally relevant cases, as done in  \cite{kolenderski_note:_2010}.
First, we will assume the spectral pump function  to be a normalized
Gaussian, namely,
\begin{equation}
\label{amplitude}
\mathcal{A}_\p^{\rm temp}(\omega_\p)=
\sqrt[4]{\frac{2\tau_\p^2}{\pi}}
e^{-\tau_\p^2(\omega_\p-\omega_{0})^2}
\end{equation}
 with $\tau_\p$ related to the pulse duration and $2 \tau_\p$ the inverse of the standard deviation of the modulus square of \eqref{amplitude}. Then, in the experimental setup we have in mind, the photons are emitted and collected using single mode fibers. We can therefore try to approximate their spatial wave functions using Gaussians (see Appendix \ref{sec:gaussian})
\begin{equation}\label{eq:spat_wave}
u_a(\kk_{a},\omega_{a})=
\sqrt{\frac{{\sf w}_{ax}{\sf w}_{ay}}{2\pi}}
\prod_{\mu=x,y}
e^{-\frac{{\sf w}_{a\mu}^2}{4}( k_{a\mu}-k_{0 a \mu})^2}
\,,
\end{equation}
where $a=\p,\ii,\s$, $k_{0 \p \mu}=0$, and $k_{0 \s \mu}$ and $k_{0 \ii \mu}$ are the central wave vectors corresponding to the directions where the photons are collected.
The parameters
${\sf w}_{a\mu}$ are the beam waists; we consider the possibility that the beams are elliptic, meaning that the beam waists ${\sf w}_{ax}$ and
${\sf w}_{ay}$ can be different.

Finally, we choose $\kk_{0\ii}$ and
$\kk_{0\s}$ such that the phase-matching and 
 energy-matching conditions are satisfied for the idler and signal photons at frequencies $\Omega_\ii$ and $\Omega_\s$, namely,
\beq
\label{phase_matching}
\begin{cases}
\Delta k_z(\kk_{0\ii},\Omega_\ii;\kk_{0\s},\Omega_\s)=0
&\text{for phase matching }\parallel \ ,
\\
\kk_{0\ii}(\Omega_\ii)+\kk_{0\s}(\Omega_\s)=0
&\text{for phase matching }\perp \ ,
\\
\Omega_\ii+\Omega_\s=\omega_0
&\text{for energy matching .}
\end{cases}
\eeq 
To solve Eq. \eqref{phase_matching}, the central wave vectors, whose direction is defined by the fiber position, can be written as
\begin{equation}\label{eq:central_vector}
\kk_{0\ii}=\frac{\omega_\ii\sin \alpha_\ii}{c}
\hat{{\bf m}}\,,\qquad
\kk_{0\s}=-
\frac{\omega_\s \sin \alpha_\s}{c}\hat{{\bf m}}\,,
\end{equation}
where $\hat{{\bf m}}$ is a unit vector in the $(x,y)$ plane and $\alpha_\ii>0$ and $\alpha_\s>0$ are the absolute values of the angles of collection direction with respect to the $z$ axis
(see Fig.\ \ref{fig:setup}). These angles are related by
$\Omega_\ii\sin \alpha_\ii=\Omega_\s\sin \alpha_\s$ to satisfy the transverse ($\perp$) phase matching condition of \eqref{phase_matching}.
The description \eqref{eq:spat_wave} holds for small angles of collection (see Appendix \ref{sec:gaussian}) so that in \eqref{eq:central_vector} we will approximate the sine function with its argument.

\section{Paraxial approximation
}\label{sec:par_appr}
\begin{figure}
     \centering
     \includegraphics[width=0.95\columnwidth]{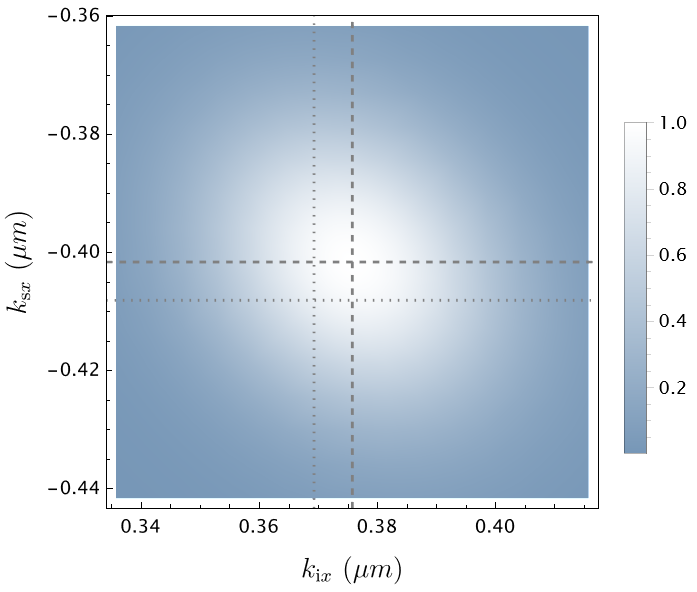}
     \caption{Normalized absolute value of the integrand in \eqref{PhiGeneral} as a function of $k_{\ii x}$ and $k_{\s x}$, for $\omega_{\ii}=\omega_{0}/2.1$ and $\omega_{\s}=\omega_{0}/1.9$. The dashed lines denote the values of $\overline{k}_{ax}$ and the dotted lines the values of $k_{a0x}$. The simulation has been performed for a type-I SPDC realized via a BBO crystal with $\lambda_0=2 \pi c/\omega_0=\SI{405}{\nano\meter}$, perfect phase matching, collection angle $\alpha \approx \SI{2.8}{\degree}$ in the $(x,z)$ plane, ${\sf w}_\ii={\sf w}_\s=\SI{50}{\micro \meter}$ and ${\sf w}_\p=\SI{25}{\micro \meter}$.}
 \label{fig:approx}
 \end{figure}

The actual computation of the integrals in \eqref{PhiGeneral} is rather difficult, due to the complicated dependence of $\Delta k_z$ on the transverse components, as is clear from \eqref{kz_general}. However,
we can note that the absolute value of the 
 integrand in \eqref{PhiGeneral} is given by the product of the spatial wavefunctions $u_a$, i.e., up to an overall factor,
 \begin{equation}\label{eq:exp}
\prod_{\mu=x,y} e^{-\frac{{\sf w}_{\ii\mu}^2}{4}(k_{\ii\mu}-k_{0\ii\mu})^2-\frac{{\sf w}_{\s\mu}^2}{4}(k_{\s\mu}-k_{0\s\mu})^2 -\frac{{\sf w}_{\p\mu}^2}{4} k_{\p \mu}^2} \ ,
\end{equation}
 with $k_{\p\mu}=k_{\ii\mu}+k_{\s\mu}$. This quantity is strongly peaked around its maximum if
\begin{equation}\label{eq:par_approx}
{\sf w}_{a\mu}\gg \lambda_a\,,\qquad a=\ii,\s,\p,
\end{equation}
a condition that is satisfied in typical SPDC sources. We can then apply the so-called paraxial approximation: A small error can be made in the computation of \eqref{PhiGeneral} by expanding $\Delta k_z$ at the second order around the maximum of \eqref{eq:exp}. For each component $\mu=x,y$, this maximum is at
\begin{equation}
    \begin{aligned}
\overline{k}_{\ii\mu}(\omega_\ii,\omega_\s)&=
k_{0 \ii\mu}(\omega_\ii)
-\frac{\overline{{\sf w}}^2_\mu}{{\sf w}_{\ii \mu}^2}
[k_{0\ii \mu}(\omega_\ii)+k_{0 \s\mu}(\omega_\s)]\\
\overline{k}_{\s\mu}(\omega_\ii,\omega_\s)&=
k_{0 \s \mu}(\omega_\s)
-\frac{\overline{{\sf w}}^2_\mu}{{\sf w}_{\s \mu}^2}
[k_{0\ii \mu}(\omega_\ii)+k_{0 \s\mu}(\omega_\s)]
\end{aligned}
\end{equation}
where we introduced the effective beam waist $\overline{{\sf w}}_\mu$:
\begin{equation}
\frac{1}{\overline {\sf w}_{\mu}^2}
=
\frac{1}{{\sf w}_{\p\mu}^2}
+
\frac{1}{{\sf w}_{\ii\mu}^2}
+
\frac{1}{{\sf w}_{\s\mu}^2} \ .
\end{equation}
Note that, although similar in spirit, our approximation around the vectors
${\overline\kk}_\ii$ and ${\overline\kk}_\s$ is different from the one discussed in \cite{kolenderski_note:_2010}, where the expansion is simply performed around the center of the two Gaussians $\kk_{0\ii}$ and $\kk_{0\s}$. Since $\overline k_{\ii\mu}(\Omega_\ii,\Omega_\s)=k_{0\ii\mu}(\Omega_\ii)$ and $\overline k_{\s\mu}(\Omega_\ii,\Omega_\s)=k_{0\s\mu}(\Omega_\s)$, the two approximations are equivalent for the central frequencies $\Omega_{\ii}$ and $\Omega_{\s}$, but the expansion around $\kk_{0\ii}$ and $\kk_{0\s}$ is not centered on the maximum of the exponential (given by $\overline{k}_{\ii \mu} $ and $\overline{k}_{\s \mu}$) for other values of $\omega_{\ii}$ and $\omega_{\s}$. 
Therefore, our choice gives a more accurate approximation of $\Phi(\omega_\ii,\omega_\s)$ with respect to the one performed in \cite{kolenderski_note:_2010} when $\omega_i\neq\Omega_i$ and $\omega_s\neq\Omega_s$. An example is given in Fig.\ \ref{fig:approx}.

To expand $\Delta k_z$ to second order, we first expand each longitudinal component $k_{az}$ as
\begin{equation}
\label{kz_expansion_text}
k_{az} 
\approx \overline{k}_{az}+\sum_\mu (k_\mu-\overline{k}_\mu){\rm K}_{1a}^\mu
+\frac12 \sum_{\mu\nu} (k_\mu-\overline{k}_\mu){\rm K}_{2a}^{\mu\nu}(k_\nu-\overline{k}_\nu) 
\end{equation}
where $\overline{k}_{az}=k_{az}(\overline{\mathbf{k}}_a)$,
$\overline{\kk}_a=(\overline k_{ax},\overline k_{ay})$ and we introduced the derivatives
\begin{equation}
{\rm K}_{1a}^{\mu}=
\left.\frac{\partial k_{az}}{\partial k_{a\mu}}
\right|_{\kk_a=\overline{\kk}_a}\,,
\qquad
{\rm K}_{2a}^{\mu\nu}=
\left.\frac{\partial^2 k_{az}}{\partial k_{a\mu}\partial k_{a\nu}}\right|_{\kk_a=\overline{\kk}_a} \ .
\end{equation}
In Eq.\ \eqref{kz_expansion_text} we also defined
\begin{equation}
    \overline k_{\p\mu}(\omega_\ii,\omega_\s)=\overline k_{\ii\mu}+\overline k_{\s\mu}
    =\left(\frac{\overline{{\sf w}}_\mu}{{\sf w}_{\p \mu}}\right)^2
(k_{0 \ii \mu}
+k_{0 \s \mu})\,.
\end{equation}

With the substitution of \eqref{kz_expansion_text}, the resulting integral in \eqref{PhiGeneral} becomes Gaussian in the transverse components, and can be solved. 
The result can be expressed in terms of some physical relevant quantities. Indeed, we can introduce the focal parameter $\xi_{a\mu}$ and
the deviation parameter $\nu_{a\mu}$, given by
\begin{equation}\label{eq:parameters}
\xi_{a\mu}(\omega_a)=-\frac{L}{{\sf w}^2_{a\mu}}{\rm K}_{2a}^{\mu \mu} \ , \quad \nu_{a\mu}(\omega_a)=-\frac{L}{2{\sf w}_{a\mu}}{\rm K}_{1a}^{\mu}
\ .
\end{equation}
As discussed in Appendix \ref{sec:gaussian}, the parameter $\nu_{\mu}$ is related
to the lateral deviation of the beam (in units of beam waist ${\sf w}_{\mu}$
inside the crystal) since, at the output of the crystal, the beam is shifted by ${\sf w}_{a\mu}\nu_{a\mu}$.
The parameter $\xi_{\mu}$, instead,
is related to the focusing condition of the beam in the $\mu$ direction \cite{Bennink_note:_2010}: If
$\xi_{\mu}\gg1$ ($\xi_{\mu}\ll1$)
the field is strongly (weakly) focused relative to the length $L$ of the crystal.

Leaving to Appendix \ref{sec:computations} a more general expression of the bi photon wave function, we report here the case in which the photons are collected in the $(x,z)$ or the $(y,z)$ plane, a very common choice in experiments. In this configuration, the expression \eqref{PhiGeneral} in the paraxial approximation becomes
\begin{equation}\label{eq:phi_fact_text}
\begin{split}
\Psi (\omega_\ii,\omega_\s)
&=
2\sqrt{2\pi} L\mathcal N
\mathcal{A}_\p^{\rm temp}(\omega_\p) \times \\
&
\times \prod_{\mu=x,y} 
\left(
\frac{\overline{{\sf w}}_\mu
e^{-\frac14 \overline{{\sf w}}_{\mu}^2( k_{0\ii \mu}+ k_{0\s \mu})^2}}{
\sqrt{{\sf w}_{\p\mu}{\sf w}_{\ii\mu}{\sf w}_{\s\mu}}} 
\right) \times \\
& \times \int_{-1}^{1} \! \! {\rm d}Z\,\,
e^{-\frac{iL\Delta \overline{k}_{z}}{2}Z}
\prod_{\mu=x,y}
\frac{\exp(-Z^2\frac{Q_\mu (Z)}{F_\mu(Z)})}{\sqrt{F_\mu(Z)}}  
\end{split}
\end{equation} 
where $\Delta \overline{k}_{z}=
\Delta k_{z}(\overline{\kk}_{\s},\omega_\s;\overline{\kk}_{\ii},\omega_\ii)$ and
\begin{align}
\label{eq:Q} Q_\mu(Z)&=A_\mu-i B_\mu Z \  , \\
\label{eq:confocal} F_\mu(Z)&=1+i \xi_\mu Z+C_\mu Z^2 \ .
\end{align}
Introducing the combination of waists and deviation parameters
\begin{equation}
    \Delta^2_{ab\mu}=\overline{{\sf w}}_\mu^2\left(\frac{\nu_{a\mu}}{{\sf w}_{b\mu}}-
 \frac{\nu_{b\mu}}{{\sf w}_{a\mu}}\right)^2 \ ,
\end{equation}
the terms in $Q_\mu$ are given by
\begin{align}
\label{eq:A} A_\mu&=
 \Delta^2_{\ii\s\mu}+
 \Delta^2_{\p\s\mu}+
 \Delta^2_{\p\ii\mu} \ , \\
 B_\mu&=
 \Delta^2_{\ii\s\mu}\xi_{\p\mu}-
 \Delta^2_{\p\s\mu}\xi_{\ii\mu}-
 \Delta^2_{\p\ii\mu}\xi_{\s\mu} \ .
 \end{align}
The expression of $F_\mu (Z)$ in \eqref{eq:confocal} instead is written in terms of  an aggregate focal parameter
\begin{equation}\label{eq:aggregrate}
    \xi_\mu =\xi_{\ii \mu}\left(1-\frac{\overline{{\sf w}}_\mu^2}{{\sf w}_{\ii \mu}^2} \right)+\xi_{\s \mu}\left(1-\frac{\overline{{\sf w}}_\mu^2}{{\sf w}_{\s \mu}^2}\right)-\xi_{\p \mu}\left(1-\frac{\overline{{\sf w}}_\mu^2}{{\sf w}_{\p \mu}^2} \right)  
\end{equation}
and the quantity
\begin{equation}
    C_{\mu}= \overline{{\sf w}}_\mu^2\left(\frac{\xi_{\s\mu} \xi_{\p\mu}}{{\sf w}_{\ii\mu}^2} +\frac{\xi_{\ii\mu} \xi_{\p\mu}}{\rm{w}_{\s\mu}^2}-\frac{\xi_{\ii\mu} \xi_{\s\mu}}{{\sf w}_{\p\mu}^2}  \right)
\end{equation}
which have already been  introduced and discussed for the collinear SPDC in \cite{Bennink_note:_2010} (Eqs.\ (14) and (15)).

Equation \eqref{eq:phi_fact_text} is the first main result of this work: It is a generalization of the expression (16) of \cite{Bennink_note:_2010}, taking into account more general angles of emission, birefringence effects
and possible transverse walk-off of the beams (see Appendix \ref{sec:gaussian} for a discussion on this point). We note that, the result obtained in Eq.\ (16) of \cite{Bennink_note:_2010} is recovered, up to different conventions in the overall factor, by noting that in the case of periodically poled crystals with collinear emission we have $Q_\mu(Z)=0$ and $\kk_{0\ii}=\kk_{0\s}=0$.
Equation \eqref{eq:phi_fact_text} is also an improvement and simplification with respect to the results contained in \cite{kolenderski_note:_2010}. The differences are in the improved choice of expansion, as already discussed, and in the factorization of the integrand in the $x$ and $y$ components, due to the particular choice of the plane of emission.

\section{Thin crystal approximation}\label{sec:thin}
For the rest of the work, we will be interested in studying the total brightness \eqref{eq:bright}, starting from some analytical results that can be obtained in the limit of thin crystals. To be considered a thin crystal, two conditions must be met: First, we require the length $L$ of the crystal to be short with respect to the confocal parameter of the  Gaussian beams corresponding to pump, idler, and signal photons. Moreover,
the lateral deviation of the beams should be negligible with respect to the beam waists. Mathematically, these conditions can be expressed in terms of the parameters in \eqref{eq:parameters} as
\begin{equation}
\label{thin_crystal}
\xi_{a\mu}\ll1 \ ,
\qquad
\nu_{a\mu}\ll1
\,,\qquad
\forall  \ a,\mu \ .
\end{equation}
At order $0$ in the expansion in $\nu_{a\mu}$ and $\xi_{a\mu}$, the $Z$ integral in \eqref{eq:phi_fact_text} is simplified into 
\begin{align}
\label{PhiThin}
\notag &\Psi(\omega_\ii,\omega_\s) = 
{4\sqrt {2\pi}}\mathcal{N} L \mathcal{A}_\p^{\rm temp}(\omega_\ii+\omega_\s) \times
\\
& \qquad \times \text{sinc}\left(\frac{L\Delta \overline{k}_{z}(\omega_\ii,\omega_\s)}{2}\right) \times \\
& \notag \qquad \qquad \times
   \prod_{\mu=x,y}\left(\frac{{\overline {\sf w}_\mu}^2}{{\sf w}_{\p\mu}{\sf w}_{\ii\mu}{\sf w}_{\s\mu}}\right)^{\frac{1}{2}}
e^{-\frac{\overline {\sf w}^2_\mu}{4} (k_{0\ii\mu}+k_{0\s\mu})^2} \ .
\end{align}
Actually, starting from the more general equation \eqref{eq:phi_gen}, it can be verified that the expression \eqref{PhiThin} holds regardless of the choice of collection plane.

 To focus on the main aspects of the problem, we can take symmetric waists in the $x$ and $y$ axes and
\begin{equation}\label{eq:waist_conf}
    {\sf w}_{\ii}={\sf w}_{\s}={\sf w} \ , \qquad {\sf w}_{\p}=r{\sf w}
\end{equation}
so that
\begin{equation}
\overline{{\sf w}}^2=\frac{r^2 {\sf w}^2}{1+2r^2}  \ .
\end{equation}
This configuration is physically relevant since the signal and idler are often collected in identical fibers, using symmetrical optical apparatuses.
Moreover, we are interested in degenerate emission $\Omega_\ii=\Omega_\s=\omega_0/2$ that implies, according to Eq.\ \eqref{phase_matching}, equal angles of collection
\begin{equation}\label{eq:conf_angle}
    \alpha_\ii=\alpha_\s \equiv \alpha \ .
\end{equation}
With this choice, by writing
\begin{equation}
\begin{split}
\label{eq:k_plane}
    \kk_{0 \s}&=(k_{0\s x},k_{0\s y})=\frac{\omega_\s \alpha}{c}(\cos \phi,\sin \phi) \ , \\
    \kk_{0\ii}&=(k_{0\ii x}, k_{0\ii y})=-\frac{\omega_\ii \alpha}{c} (\cos \phi,\sin \phi) 
\end{split}
\end{equation}
we have
\begin{equation}
    \sum_{\mu=x,y} \frac{\overline{{\sf w}}^2}{4}(k_{0 \ii \mu}+k_{0\s \mu})^2=\frac{\overline{{\sf w}}^2\alpha^2}{4c^2}(\omega_\s-\omega_\ii)^2 \ .
\end{equation}
Nonetheless, the following discussion can be generalized to arbitrary parameters.

\subsection{Longitudinal  perfect phase matching}
Our main goal is to find the optimal focusing conditions in order to have a bright source: in the configuration \eqref{eq:waist_conf} this corresponds to finding the optimal values of ${\sf w}$ and $r$ to maximize \eqref{eq:bright}. To begin with, we can consider this problem when the longitudinal phase mismatch
satisfies $L\Delta \overline{k}_{z} \ll 1$
for all the values of $\omega_\ii$ and $\omega_\s$ that do not suppress the two Gaussian terms in \eqref{PhiThin}.
In this case, which we will call longitudinal perfect phase matching, the sinc term is approximately equal to 1 and the constraints to the spectral properties of $\Psi$ are due to the coherence length $\tau_\p$ of the pump, the central frequency $\omega_0$ of the pump, the collection angle $\alpha$, and the beam waists. 
This condition is clearly more reasonable in the presence of short crystals ($L$ small) and when there is a good spatial overlap between the photons inside the crystal. Also, it is better satisfied in the presence of spectral filters, but we will consider this configuration in Sec.\ \ref{sec:filters}. 

In this section the only restriction to the collected frequencies is given by the finite transmission range of the crystal used in the SPDC process.
In a first approximation, we can model the transmission range with a step function: Calling $\omega_b$ and $\omega_t$ the minimum and the maximal frequencies transmitted in a material, we impose the conditions
\begin{equation}\label{eq:bound}
   \omega_b \le \omega_\ii, \omega_\s, \omega_\p \le \omega_t 
\end{equation}
and we recall that $\omega_\p=\omega_\ii+\omega_\s$. These constraints restrict the integration domain in \eqref{eq:bright} and, in longitudinal perfect phase matching, we obtain
\begin{equation}\label{eq:bright_int_0}
  R_{\rm tot}\propto  \hspace{-0.5cm}\int\limits_{\scriptscriptstyle\mspace{-12mu}
             \begin{subarray}{l}
            \omega_b \le \omega_\ii, \omega_\s, \omega_\ii+\omega_\s \le \omega_t
             \end{subarray}\mspace{-24mu}} \! \! \! {\rm d} \omega_\ii {\rm d} \omega_\s \  e^{-2\tau_p^2(\omega_\ii+\omega_\s-\omega_0)^2-\frac{\overline{{\sf w}}^2}{2c^2}\alpha^2(\omega_\ii-\omega_\s)^2}\ .
\end{equation}
After performing the change of variables $u=\omega_\ii+\omega_\s$ and $v=\omega_\ii-\omega_\s$ and solving the integral in $v$, we arrive at
\begin{equation}
    \begin{split}\label{eq:bright_int}
&R_{\rm tot} \! = \! {32 \pi}\mathcal{N}^2 L^2 \frac{\tau_\p c{\overline {\sf w}}^3}{\alpha r^2 {\sf w}^6 }\times \\ &\times \int_{2 \omega_b}^{\omega_t}{\rm d}u \  e^{-2\tau_\p^2(u-\omega_0)^2} \text{erf}\left(\frac{\overline{{\sf w}}\alpha}{\sqrt{2}c}(u-2\omega_b) \right) 
 \end{split}
\end{equation}
where erf denotes the error function and we restored the overall coefficients.
 We are interested in small angles $\alpha$, so it is useful to define $x=u-2 \omega_b$ and consider the Maclaurin series
\begin{equation}\label{eq:maclaurin}
    \text{erf}\left(\frac{\overline{{\sf w}}\alpha}{\sqrt{2}c}x \right)=\frac{2}{\sqrt{\pi}}\sum_{n=0}^{\infty}\frac{(-1)^n}{n!(2n+1)}\left(\frac{\overline{{\sf w}}\alpha}{\sqrt{2}c}x \right)^{2n+1} \ .
\end{equation}
Plugging this expansion in \eqref{eq:bright_int} and inverting the order of sum and integral, we can write the total brightness as
\begin{equation}
\label{eq:bright_final}
\begin{split}
  R_{\rm tot}=&32 L^2 \mathcal{N}^2 \frac{r^2\sqrt{2 \pi}}{(1+2 r^2)^2{\sf w}^2\tau_\p} \times \\
  &\times\sum_{n=0}^\infty \frac{(-1)^n}{2^n (1+2n)n!}\left( \frac{r^2}{1+2r^2}\right)^n \left( \frac{\alpha {\sf w}}{\tau_\p  c}\right)^{2n}d_n \ ,
\end{split}
\end{equation}
where we introduced 
\begin{equation}\label{eq:dn}
\begin{split}
    d_n= \int_0^{\tau_\p(\omega_t-2\omega_b)} {\rm d}y \ e^{-2(y+\tau_\p(2\omega_b- \omega_0))^2} y^{1+2n} \ . 
\end{split}    
\end{equation}
The integral defining $d_n$ can be solved for each $n$, but its expression is not particularly useful for our discussion. Instead, we are more interested in understanding some limiting situations in which the series \eqref{eq:maclaurin} (and consequently \eqref{eq:bright_final}) can be truncated or simplified. The relevant quantity is the argument of the $\text{erf}$ in the central value $\omega_0$, namely, the adimensional parameter $\alpha{\sf \overline{w}}(\omega_0-2 \omega_b)/c$.

\subsubsection{Collinear case}
When $\alpha=0$ or more generally the quantity $\alpha{\sf \overline{w}}(\omega_0-2 \omega_b)$ is very small compared with $c$, the series \eqref{eq:maclaurin} can be truncated at order 0. In this case, the brightness becomes
\begin{equation}\label{eq:bright_coll}
  R_{\rm tot}=32 L^2 \mathcal{N}^2 \frac{r^2 \sqrt{2 \pi}}{(1+2 r^2)^2{\sf w}^2\tau_\p}d_0 \ .
\end{equation}
The maximum of \eqref{eq:bright_coll} is clearly at
\begin{equation}
    r^*=\frac{1}{\sqrt{2}} \ .
\end{equation}
This optimal condition emerges in the discussions of various previous works  \cite{Bennink_note:_2010,Ling2008,Smirr2013} studying collinear emission or assuming perfect transverse phase matching. Indeed, the latter condition corresponds to not having the second term in the exponential of \eqref{eq:bright_int_0}, as in the collinear case. Note that the brightness \eqref{eq:bright_coll} increases by decreasing ${\sf w}$, but the divergence in ${\sf w}=0$ is not physical since we cannot apply the thin crystal approximation
for small values of ${\sf w}$ and, before that, we cannot apply the paraxial approximation for waists of the same order of the wavelengths.

\subsubsection{Large emission angle} Another interesting limit is the one in which  $\alpha\overline{{\sf w}}(\omega_0-2 \omega_b)$ is much larger than $c$. In this case the integral in \eqref{eq:bright_int} can be solved by replacing the $\text{erf}$ function with its asymptotic value $1$. Note that this substitution introduces an error of less than $1 \% $ for arguments $x \gtrsim 2 \sqrt{2}c/\overline{{\sf w}}\alpha$. The resulting brightness is
\begin{equation}
\begin{split}
\label{eq:bright_non_coll}
  R_{\rm tot}&=8\sqrt{2} L^2 \mathcal{N}^2 c \frac{\pi^{3/2}r }{(1+2 r^2)^{3/2}{\sf w}^3\alpha}\times\\ &\times \bigg(\text{erf}\left(\sqrt{2}\tau_\p (\omega_0-2 \omega_b)\right)-\text{erf}\left(\sqrt{2}\tau_\p (\omega_0-\omega_t) \right) \bigg)
\end{split}
\end{equation}
with a maximum at
\begin{equation}
    r^*=\frac{1}{2} \ .
\end{equation}
The expression \eqref{eq:bright_non_coll} was also derived in \cite{kolenderski_note:_2010} for the case of large $\tau_\p \omega_0$ and integrating over the whole plane of the frequencies $(\omega_\ii,\omega_\s)$. In fact, the integral in \eqref{eq:bright_int_0} is made of two exponential terms which
select, in the plane of integration $(\omega_\ii,\omega_\s)$, a contribution from a strip around the line $\omega_\s=-\omega_\ii+\omega_0$ and another around the line $\omega_\ii=\omega_\s$. The “width" of the strips depends on the values of $\tau_\p$ and $\alpha {\sf w}$, respectively. The intersection of the two strips gives the domain which is relevant for the integration. As $\alpha \sf w$ and $\tau_\p$ increase, this domain becomes smaller  and smaller and we can expand the extrema of integration in the integral \eqref{eq:bright_int_0} to $(-\infty,\infty)$, as discussed in \cite{Dragan2004}. In this way, the integration can be performed via standard techniques recovering the result of \cite{kolenderski_note:_2010}. We stress, however, that \eqref{eq:bright_non_coll} holds only when $\alpha {\sf w}(\omega_0-2 \omega_b)\gg c$. For this reason, the divergence at $\alpha {\sf w}=0$ in \eqref{eq:bright_non_coll} is unphysical.

\paragraph{Optimal angle in the $(x,y)$ plane}
In the case of sufficiently large emission angles, we can ask whether the rate emission is constant in the plane $(x,y)$.
For equal waists ${\sf w}_{ax}={\sf w}_{ay}$, we just found that there is no dependence on the direction in the $(x,y)$ plane. However, we can try to generalize the previous result as follows. Let us relax, only in this paragraph, the conditions \eqref{eq:waist_conf} and assume different waists on the $x$ and $y$ directions and between the idler and signal. Integrating the frequencies over the whole real axes, the brightness is
\begin{equation}
\label{Rthinsimple}
R_{\rm tot}\propto
\frac{1}{\sqrt{\overline {\sf w}_x^2\cos^2 \phi+\overline {\sf w}_y^2\sin^2\phi}}\prod_{\mu=x,y }\frac{\overline {\sf w}_\mu^2}{
{\sf w}_{\p \mu}{\sf w}_{\ii \mu}
{\sf w}_{\s \mu}
}\, ,
\end{equation}
where we used the definitions in \eqref{eq:k_plane}.
Assuming the waists ${\sf w}_{\p\mu}$, ${\sf w}_{\ii\mu}$, and ${\sf w}_{\s\mu}$ are fixed, we can find the best emission angle by minimizing the square root in \eqref{Rthinsimple}, which gives ($n \in \mathds{Z}$)
\begin{equation}
    \begin{cases}
    \phi= (2n+1)\frac{\pi}{2} \ , \qquad &\text{if} \ \overline{{\sf w}}_x > \overline{{\sf w}}_y \\
      \phi= n \pi  \ , \qquad &\text{if} \ \overline{{\sf w}}_x < \overline{{\sf w}}_y \\
    \end{cases}
\end{equation}
while, if $\overline{{\sf w}}_x=\overline{{\sf w}}_y$ there is not a privileged angle of emission, as found before. 
Therefore, the optimal emission plane is related to the direction of the maximal value of the effective beam waist.

\begin{figure}[t]
    \centering
    \includegraphics[width=1\columnwidth]{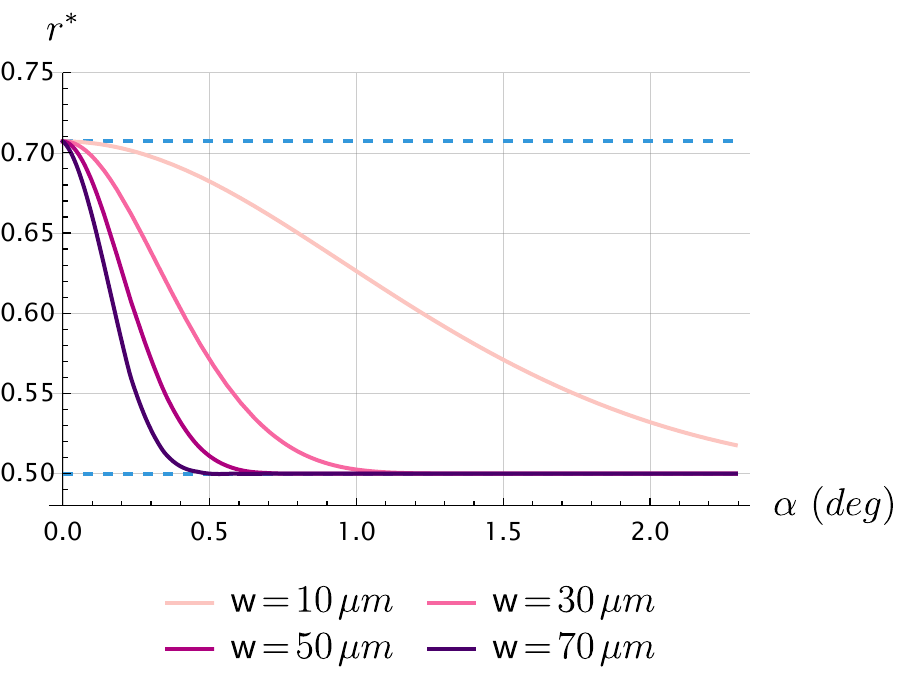}
    \caption{Value of the optimal $r$ that maximizes \eqref{eq:bright_int} as a function of $\alpha$. The optimal $r$ changes in the region $r \in [\frac{1}{2},\frac{1}{\sqrt{2}}]$.} 
    \label{fig:numerical}
\end{figure}

\subsubsection{Intermediate values} For intermediate values of $\alpha{\sf w}(\omega_0-2 \omega_b)$, it is natural to expect a smooth transition between the values $r^*=1/\sqrt{2}$ and $r^*=1/2$.
We checked this behavior numerically  by maximizing the expression in \eqref{eq:bright_int}. The results are shown in Fig.\ \ref{fig:numerical} where the simulations have been performed for three different values of the waist ${\sf w}$, $\tau_\p=\SI{e2}{\femto \second}$ and with a free space central pump wavelength $\lambda_0=\SI{405}{\nano \meter}$.  The domain of integration has been chosen between the frequencies corresponding to $\lambda_b=\SI{0.2}{\nano \meter}$ and $\lambda_t=\SI{2.2}{\nano \meter}$, i.e.\ in the clear transmission range of a BBO crystal, discussed in \cite{sellmeier}.

\begin{figure}[t]
    \centering
    \includegraphics[width=1\columnwidth]{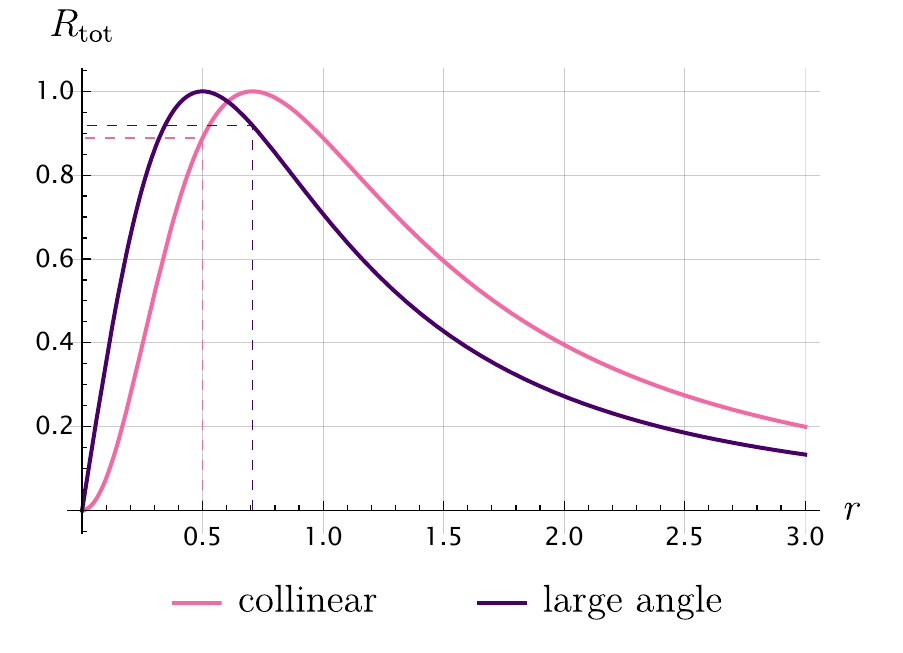}
    \caption{Variation of the (normalized to its maximum) brightness as a function of the ratio $r$. The  collinear plot was obtained from Eq.\ \eqref{eq:bright_coll}. The large-angle curve was obtained in the $\alpha {\sf w}(\omega_0-2 \omega_b)\gg c$ limit, see Eq.\ \eqref{eq:bright_non_coll}. The dotted lines are at the values $r=1/2$ and $r=1/\sqrt{2}$.}
    \label{fig:optimal}
\end{figure}

In conclusion, we have shown how the waist ratio $r={\sf w}_\p/{\sf w}_\s={\sf w}_\p/{\sf w}_\ii$ can be chosen to optimize the source brightness. The value $r^*=1/\sqrt{2}$ is reported in the literature as
the optimal value in the thin-crystal approximation  and for collinear emission \cite{Ling2008,Smirr2013}, without reference to different values of the emission cone aperture.
Here we have demonstrated instead that the optimal $r$ depends on the emission angle, as shown in Fig.\ \ref{fig:numerical}. For a large aperture, we obtained 
$r^*=1/2$, a value already derived in \cite{kolenderski_note:_2010} without the crucial observation that it holds only in the limit
of a large emission aperture angle. Note that the right choice of the optimal ratio is quite relevant. As shown in Fig.\ \ref{fig:optimal}, by choosing $r=1/2$ in the collinear case, instead of the optimal value $r=1/\sqrt{2}$, there is a loss of brightness of around $10 \%$. A similar situation holds in the non-collinear case, inverting the two values.

\subsection{Spectral filters}\label{sec:filters}
In many experimental and technological implementations, only the photons in a small wavelength range are desired and spectral filters are inserted before collecting the signal and idler. To reproduce this setup, we assume that the filters act as step functions and impose on the idler and signal frequencies the restrictions
\begin{equation}\label{eq:dom_filters}
    \frac{\omega_0}{2}-\delta \le \omega_\ii,\omega_\s \le \frac{\omega_0}{2}+\delta  
\end{equation}
where $2\delta$ denotes the frequency width selected by the filters. The corresponding brightness is
\begin{equation}
\begin{split}
R_{\rm tot} \! = \! {32 \pi}\mathcal{N}^2 L^2 \frac{\tau_\p c{\overline {\sf w}}^3}{\alpha r^2{\sf w}^6 }\hspace{-0.08cm}\int_{-2 \delta}^{2\delta} \hspace{-0.2cm}{\rm d}u \hspace{0.03cm} e^{-2\tau_\p^2 u^2} \text{erf}\left(\frac{\overline{{\sf w}}\alpha}{\sqrt{2}c}(2 \delta-|u|) \right) 
 \end{split}
\end{equation}
where, starting from \eqref{eq:bright_int} with the constraints \eqref{eq:dom_filters}, we performed the change of variables $u=\omega_\ii+\omega_\s-\omega_0$ and $v=\omega_\ii-\omega_\s$ and we solved the integral in $v$. For sufficiently narrow filters $\delta \ll c/{\sf w}\alpha$, the argument  of the erf function is small (since the integration variable $u$ is bounded by $\lvert 2 \delta \lvert$) and we can expand the erf function around $0$. Note that for a collection angle of $2^{\circ}$, a waist ${\sf w}=\SI{30}{\micro \meter}$ and a central wavelength $\lambda=\SI{810}{\nano \meter}$, the condition of a narrow filter selects wavelengths much smaller than about $\SI{100}{\nano \meter}$, a very common configuration. Performing the expansion to the third order in $\delta$ and solving the integral, we find
\begin{equation}\label{eq:R_filter}
  R_{\rm tot}\approx {128}\mathcal{N}^2 \frac{\sqrt{2\pi} L^2\tau_\p \delta^2 r^2}{{\sf w}^2 (1+2r^2)^2}
  \left[1-\frac{\alpha^2{\sf w}^2\delta^2r^2}{3c^2(1+2r^2)}\right]\,.
\end{equation}
Hence, in the presence of narrow filters, for which only the first term in parentheses \eqref{eq:R_filter} can be considered,
the brightness does not depend on the angle $\alpha$, at least in the thin-crystal limit, and its maximum is at $r^*=1/\sqrt{2}$. Note also that the perfect-phase-matching approximation is well satisfied, since we are only selecting a small range of frequencies around $(\Omega_\ii,\Omega_\s)$ for which $\Delta k_z(\Omega_\ii,\Omega_\s)=0$.

\subsection{The sinc contribution}\label{sec:sinc}
\begin{figure}[t]
    \centering
    \includegraphics[width=1\columnwidth]{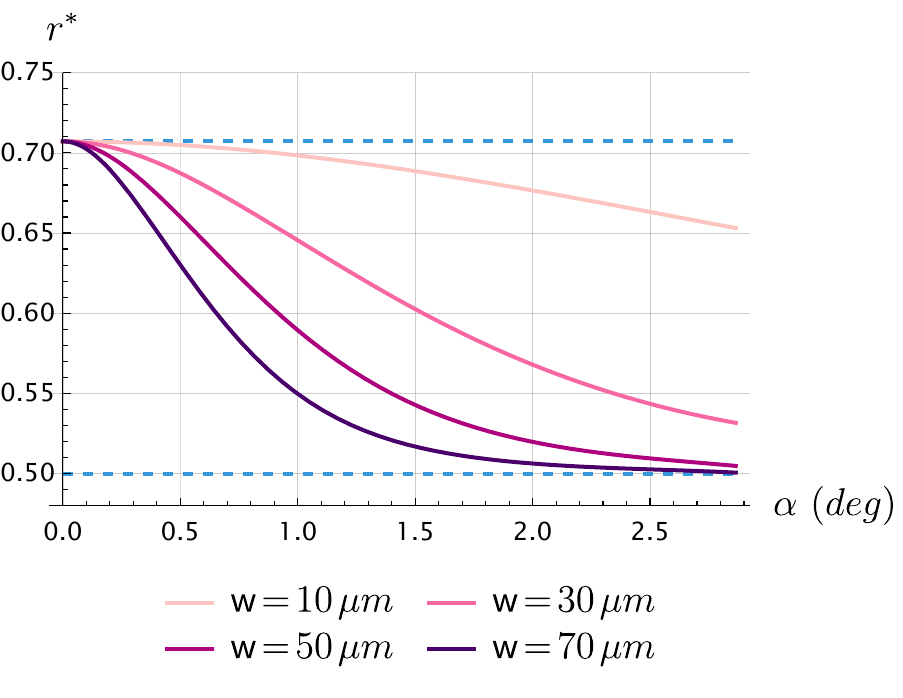}
    \caption{Optimal $r$ to maximize the thin limit brightness including the sinc contribution, as in \eqref{PhiThin}. The crystal length $L=\SI{100}{\micro \meter}$ and emission is in the $(x,z)$ plane.}
    \label{fig:numsinc}
\end{figure}

The last missing ingredient in our thin limit discussion is the sinc contribution in \eqref{PhiThin}, which we will only introduce numerically. 
Figure \ref{fig:numsinc} shows a plot of the optimal value of $r$, obtained using the thin limit bi-photon wave function \eqref{PhiThin} which includes the sinc contribution. 
The simulations have been performed for type I SPDC, $e \to o+o$, with degenerate emission and central pump wavelength at $\lambda_0=\SI{405}{\nano \meter}$. The pump pulse duration has been fixed at $\tau_\p=\SI{e2}{\femto \second}$. In order to satisfy the phase matching conditions, we considered a BBO crystal and used the Sellmeier equations in \cite{sellmeier}. 
For each value of $\alpha$, we computed the optimal angle $\theta$ between pump and optical axis to satisfy $\Delta k_z=0$ at the central frequencies $\omega_0,\Omega_\ii$, and $\Omega_\s$. By doing this we ensured that any variation in the emission rate is due only to geometrical factors and not to phase mismatch. The domain of integration has been chosen to be equal to the transmission range discussed in \cite{sellmeier}.
With respect to the perfect-phase-matching case of Fig.\ \ref{fig:numerical}, the introduction of the sinc term produces curves with a smaller slope, without changing the extreme values.

\section{Beyond the thin crystal limit}\label{sec:numerical}
So far we have discussed the brightness optimal conditions in the thin-crystal limit, trying to understand the role of different experimental setups and computational assumptions. In this last section we go beyond the thin limit and report some numerical simulations obtained from the full expression of the bi photon wave function, given in Eq.\ \eqref{eq:phi_fact_text}.
  Our aim is to maximize the brightness in terms of the ratio $r$ and the emitted photons waist ${\sf w}$ for different values of collection angle $\alpha$ and crystal length.

\subsection{Setup}

The simulations have been performed for degenerate type-I SPDC, $e \to o+o$, in the same setup described in Sec.\ \ref{sec:sinc}. We assumed degenerate emission in the $(x,z)$ plane with symmetric angles of collection as in \eqref{eq:conf_angle} and the waist configuration of \eqref{eq:waist_conf}. Simulations in the $(y,z)$ plane do not present significant differences.
We considered two different values for the length of the crystal, $L=100$ and $L=\SI{500}{\micro \meter}$, and four different values for the signal/idler waist ${\sf w}$, from ${\sf w}=10$ to ${\sf w}=\SI{70}{\micro \meter}$.

To understand the validity of the thin-crystal approximation in these cases, it is convenient to evaluate $A_\mu$ \eqref{eq:A}, which is a combination of lateral deviations and waists, and the aggregate confocal parameter $\xi_\mu$ \eqref{eq:aggregrate}. When these two quantities are very small, we can use the thin crystal approximation \eqref{thin_crystal}.
The approximate values of $A_\mu$ and $\xi_\mu$ for our configurations are reported in Tables \ref{tab:100} and \ref{tab:500}. Each quantity has been evaluated at the optimal ratio $r^*$.

\begin{table}[b]
        \centering
        \begin{tabular}{ccccccccc}
\toprule
 ${\sf w}$ & \multicolumn{2}{c}{$\SI{10}{\micro \meter}$}& \multicolumn{2}{c}{$\SI{30}{\micro \meter}$} & \multicolumn{2}{c}{$\SI{50}{\micro \meter}$} & \multicolumn{2}{c}{$\SI{70}{\micro \meter}$} \\
  $\alpha$ & $\SI{0}{\degree}$ &  $\SI{2.5}{\degree}$ & $\SI{0}{\degree}$ &  $\SI{2.5}{\degree}$ & $\SI{0}{\degree}$ &  $\SI{2.5}{\degree}$ & $\SI{0}{\degree}$ &  $\SI{2.5}{\degree}$ \\
  \toprule
     $ \xi_{x}$  & \multicolumn{2}{c}{\multirow{2}{*}{$0.07$}} & \multicolumn{2}{c}{\multirow{2}{*}{$0.008$}} & \multicolumn{2}{c}{\multirow{2}{*}{$0.003$}} & \multicolumn{2}{c}{\multirow{2}{*}{$0.001$}} \\
     $ \xi_{y}$  &  \\
    \midrule
     $A_{x}$  & 0 & $0.03$ & 0 & $0.004$ &0 & $0.001$ &0 & $7 \cdot 10^{-4}$ \\
     $A_{y}$ & $0.11$ & $0.12$ & $0.01$ & $0.02$& $0.005$ &  $0.006$ & $0.002$ &  $0.003$ \\
\bottomrule
\end{tabular}
       \caption{Approximate values of $\xi_\mu$ and $A_\mu$ for our configurations for $L=\SI{100}{\micro \meter}$}
       \label{tab:100}
    \end{table}

    \begin{table}[b]
        \centering
        \begin{tabular}{ccccccccc}
\toprule
 ${\sf w}$ & \multicolumn{2}{c}{$\SI{10}{\micro \meter}$}& \multicolumn{2}{c}{$\SI{30}{\micro \meter}$} & \multicolumn{2}{c}{$\SI{50}{\micro \meter}$} & \multicolumn{2}{c}{$\SI{70}{\micro \meter}$} \\
  $\alpha$ & $\SI{0}{\degree}$ &  $\SI{2.5}{\degree}$ & $\SI{0}{\degree}$ &  $\SI{2.5}{\degree}$ & $\SI{0}{\degree}$ &  $\SI{2.5}{\degree}$ & $\SI{0}{\degree}$ &  $\SI{2.5}{\degree}$ \\
  \toprule
     $ \xi_{x}$  & \multicolumn{2}{c}{\multirow{2}{*}{$0.38$}} & \multicolumn{2}{c}{\multirow{2}{*}{$0.04$}} & \multicolumn{2}{c}{\multirow{2}{*}{$0.01$}} & \multicolumn{2}{c}{\multirow{2}{*}{$0.007$}} \\
     $ \xi_{y}$  &  \\
    \midrule
     $A_{x}$  & 0 & $0.86$  & 0 & $0.10$ &0 & $0.03$ &0 & $0.02$ \\
     $A_{y}$  & $1.63$ & $1.88$ & $0.28$ & $0.32$& $0.11$ &  $0.13$ & $0.06$ &  $0.08$ \\
\bottomrule
\end{tabular}
        \caption{Approximate values of $\xi_\mu$ and $A_\mu$ for our configurations for $L=\SI{500}{\micro \meter}$}
        \label{tab:500}
     \end{table}

\subsection{Optimal ratio $r$}
\begin{figure}
  \begin{tabular}{@{}c@{}}
	\includegraphics[width=1\columnwidth]{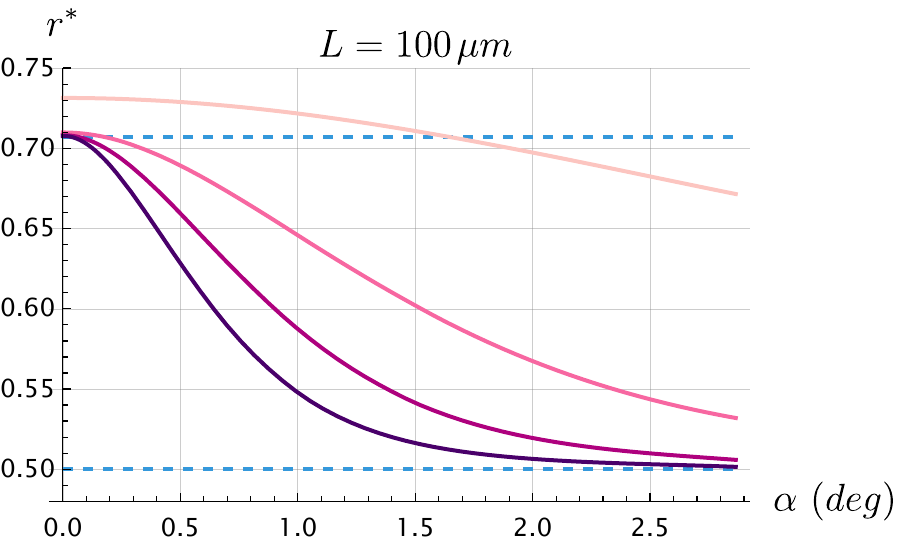}
  \end{tabular}
  \vspace{\floatsep}
  \begin{tabular}{@{}c@{}}
 \includegraphics[width=1\columnwidth]{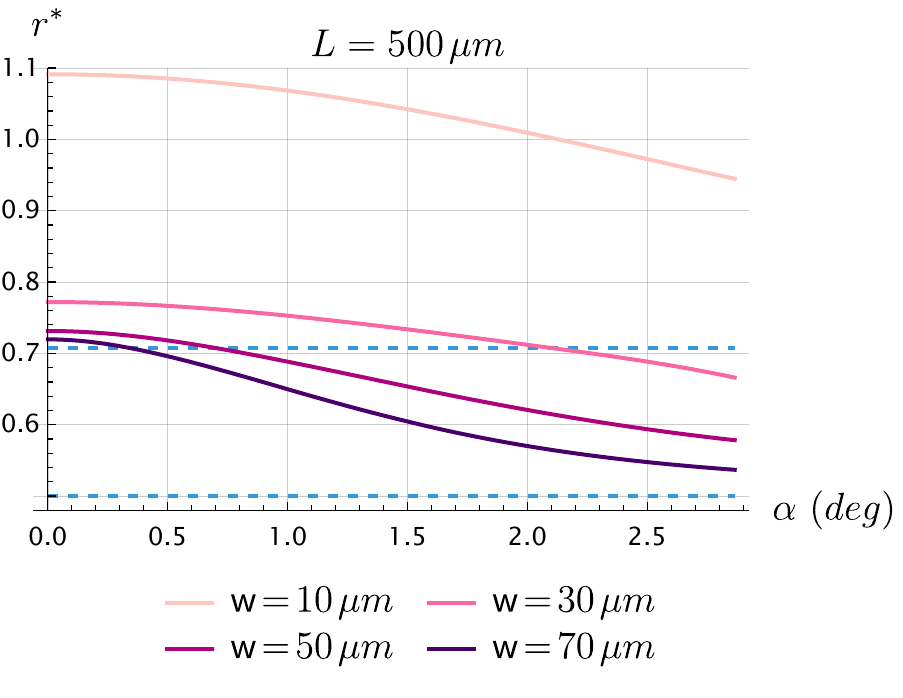}
\end{tabular}
	\caption{Transition between different optimal values moving from the collinear case to larger collection angles, at different focusing conditions. The dashed lines are at the optimal values $r^*=1/\sqrt{2}$ and $r^*=1/2$. Note the different scales between the two plots.}
	\label{fig:num_full}
\end{figure}

The graphs in Fig.\ \ref{fig:num_full} represent the optimal ratios to maximize the brightness, varying the angle of collection $\alpha$.
Substantial deviations from the thin-limit discussion emerge for all the configurations at
$L=\SI{500}{\micro \meter}$ and for the (rather strong) focusing condition ${\sf w}=\SI{10}{\micro \meter}$ at $L=\SI{100}{\micro \meter}$. In all these cases, the optimal points are subjected to a shift towards larger values, compared to those found in the thin limit. 
At the same time, looking at Tables \ref{tab:100} and \ref{tab:500}, we note that the values of $A_\mu$, which are related to the transverse walk-off, are significantly larger than the values of $\xi_\mu$. It is then reasonable to suppose that the shift is mainly due to the transverse walk-off of the beams, which becomes important for longer crystals. To better justify this claim, consider the expression  of the bi-photon wavefunction in Eq.\ \eqref{eq:phi_fact_text}. Performing an expansion at the first order in $\xi_{\mu}$, we can write
\begin{equation}\label{eq:expansions_xi}
\begin{aligned}
\frac{1}{\sqrt{F_\mu}}& \approx 1-\frac{i}{2}\xi_\mu Z  \ , \\
   e^{-Z^2\frac{Q_\mu}{F_\mu}}& \approx e^{-Z^2 A_\mu}(1+i Z^3(B_\mu+\xi_\mu A_\mu)) \ .
\end{aligned}
\end{equation}
Assuming the longitudinal perfect-phase-matching approximation ${L \Delta k_{z}\ll 1}$ and substituting the expansions \eqref{eq:expansions_xi}, it is easy to solve the $Z$-integral in \eqref{eq:phi_fact_text} to find
\begin{align}
\label{eq:thin_orders}
\notag &\Psi(\omega_\ii,\omega_\s) \approx
{2\pi\sqrt {2}}\mathcal{N} L\mathcal{A}_\p^{\rm temp}(\omega_\ii+\omega_\s)e^{-\frac{\overline {\sf w}^2}{2} (k_{0\ii\mu}+k_{0\s\mu})^2}\times
\\
& \qquad \qquad \times \frac{\text{erf}\left(\sqrt{A_x+A_y}\right)}{\sqrt{A_x+A_y}} \left(\frac{{\overline {\sf w}}^2}{{\sf w}_{\p}{\sf w}_{\ii}{\sf w}_{\s}}\right)
 \ .
\end{align}
Note that in this expression there is no dependence on $\xi_\mu$, but only on the deviation parameters $\nu_{a\mu}$, which are implicit in $A_\mu$. We now focus on the collinear emission with $\alpha=0$. In this configuration, $A_x=0$ while
\begin{equation}
A_y=\frac{L^2}{2 {\sf w}^2}\frac{1}{1+2r^2}\beta_\p^2
\end{equation}
with
\begin{equation}
    \beta_\p=n_\theta^2\left(\frac{1}{n_e^2}-\frac{1}{n_o^2}\right)\sin\theta\cos\theta \ ,
\end{equation}
where $\theta$ is the angle between the optical axis of the crystal and the propagation direction of the pump.
The refraction indices in the previous expression are frequency dependent and evaluated in $\omega_\p=\omega_\ii+\omega_\s$. However, in the first approximation, we can assume them to be constant and
evaluate them in the fixed central frequency $\omega_\p=\omega_0$. With this assumption, the collinear brightness can be derived as in \eqref{eq:bright_coll}:
\begin{equation}\label{eq:bright_coll_corr}
  R_{\rm tot}\hspace{-0.1cm}\approx \hspace{-0.1cm}  \frac{16\sqrt{2} \pi^{3/2} \mathcal{N}^2 d_0 }{\beta_\p^2\tau_\p}\frac{r^2}{(1+2 r^2)}\text{erf}\left(\frac{L \beta_\p\lvert_{\omega_\p=\omega_0}}{\sqrt{2}{\sf w}(1+2r^2)^{1/2}} \right)^2 .
\end{equation}

\begin{figure}
  \begin{tabular}{@{}c@{}}
\includegraphics[width=1\columnwidth]{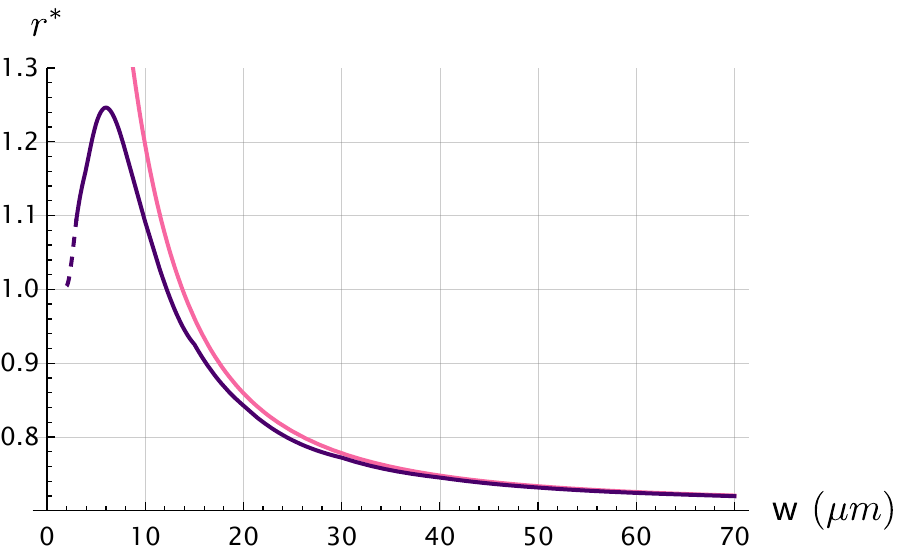}
  \end{tabular}
	\caption{Optimal ratio $r$ varying the waist ${\sf w}$, for collinear emission, with $L=\SI{500}{\micro \meter}$. The purple (darker) line has been obtained starting from the wavefunction \eqref{eq:phi_fact_text} and the pink (lighter) line by maximizing \eqref{eq:bright_coll_corr}.}
	\label{fig:expanded}
\end{figure}

The maximization of \eqref{eq:bright_coll_corr} with respect to $r$ gives the pink plot in Fig.\ \ref{fig:expanded}, where instead the purple curve represents the values obtained from the full bi-photon wave function \eqref{eq:phi_fact_text}, without the expansion for small $\xi_\mu$. 
For sufficiently large waists, the brightness \eqref{eq:bright_coll_corr} reproduces well the shifting behaviour of the optimal ratio.
Since \eqref{eq:bright_coll_corr} depends on the deviation parameter $\nu_{a y}$, and not on the confocal parameter $\xi_\mu$, we can deduce that the shifting is mostly due to the transverse walk-off, at least in regimes far from strong focusing conditions. From a physical point of view, this result can be easily interpreted: the SPDC process can only occur if there is good overlap between the three beams in the process. The increase of the pump waist compensates for the loss of overlap due to the walk-off. Finally, we note that the importance of walk-off effects has been variously discussed in the literature \cite{Dragan2004,Ling2008,BOVINO2003343, Kurtsiefer}.

\subsubsection{Other SPDC processes}

The above simulations and analysis were performed for type-I SPDC. 
Nevertheless, it is natural to expect 
a variation of the optimal ratio, similar to that just discussed, in all processes where transverse walk-off is present and the crystals are sufficiently long.

An important and particularly bright setup is the one in which the SPDC is that in which the SPDC is realized using periodically poled crystals with collinear emission and extraordinary waves propagating along one of the principal axes. In this case there is no transverse walk-off and, in light of the previous discussion, the natural expectation for this configuration is to have $r^*=1/\sqrt{2}$ for degenerate emission. This is in agreement with the results of \cite{Bennink_note:_2010}.

\subsection{Optimal waist ${\sf w}$}
\begin{figure}
  \begin{tabular}{@{}c@{}}
	\includegraphics[width=1\columnwidth]{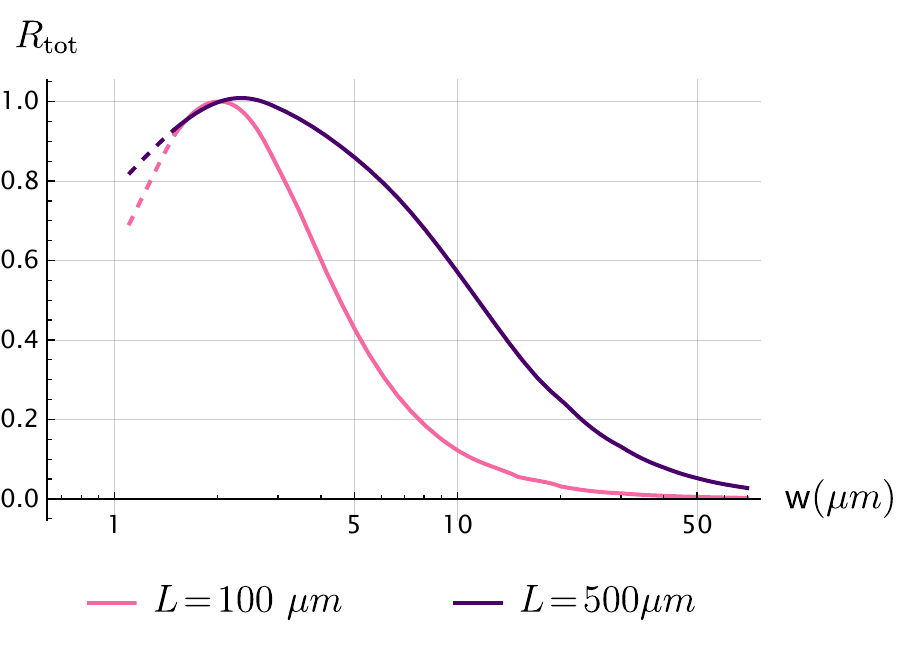}
  \end{tabular}
	\caption{Total brightness normalized to its maximum as a function of the waist ${\sf w}$, with $L=\SI{500}{\micro \meter}$, in the collinear case. The curves have been produced by interpolating a finite number of waist configurations. For each waist ${\sf w}$ we first found the optimal value of $r$ and then computed the brightness.}
	\label{fig:absolute}
\end{figure}

All the expressions for the brightness derived so far have their maximum when ${\sf w} \to 0$. As already discussed, however, for very small ${\sf w}$ the thin crystal limit or the expansion in $\xi$ of the preceding section are not justified. For this reason, we need to maximize the brightness derived from the full expression \eqref{eq:phi_fact_text}. Applying a standard numerical method we get the two plots in Fig.\ \ref{fig:absolute}, with maxima corresponding to values of ${\sf w}$ of few micrometers. 
Nonetheless, the exact value of the optimal ${\sf w}$ cannot be satisfactorily obtained from our discussion: At the very beginning of our treatment we used the paraxial approximation \eqref{eq:par_approx}, which is not true for extreme focusing conditions. Thus, what we can actually learn is that, at least for the lengths of the crystals considered in this work, we expect to increment the brightness by decreasing the idler/signal waist ${\sf w}$ to a value of a few micrometers.

\subsection{Collinear versus non-collinear brightness} 
\begin{figure}
  \begin{tabular}{@{}c@{}}
	\includegraphics[width=1\columnwidth]{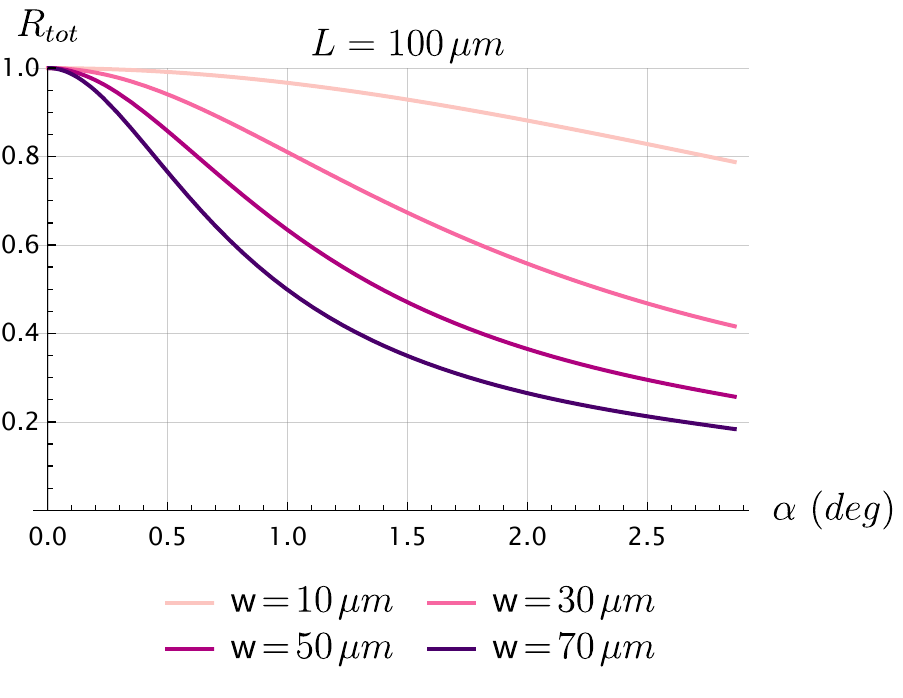}
  \end{tabular}
	\caption{Normalized total brightness as a function of the emission angle, with $L=\SI{100}{\micro \meter}$. For each waist ${\sf w}$ and angle $\alpha$ we first found the optimal value of $r$ and then computed the brightness.}
	\label{fig:coll_vs_noncoll}
\end{figure}
Finally, we can study the value of the absolute brightness as a function of the emission angle, for a given idler/signal waist. From the exponential terms in the expression \eqref{eq:phi_fact_text}, it is natural to expect the brightness to decrease for larger collection angles and larger waists ${\sf w}$. An example of this behavior, obtained with the same numerical simulations of the previous sections, is reported in Fig.\ \ref{fig:coll_vs_noncoll} for the case $L=\SI{100}{\micro \meter}$.

\section{Conclusions}\label{sec:conclusion}
In this work we studied degenerate SPDC emission, when the photons in the process are emitted and collected by single mode fibers. 
We clarified the relation between different optimal focusing conditions proposed in the literature to maximize the total brightness. In the thin-crystal limit, we gave an analytical derivation of various results, considering the role of collinear and non-collinear emission, together with the finite transmission range of the crystal and the presence of spectral filters. We found that the brightness can be increased by decreasing the signal/idler waists ${\sf w}$, assumed to be equal, whereas the optimal ratio $r$ between the pump waist and ${\sf w}$ is $1/\sqrt{2}$ for collinear emission and tends to $1/2$ for larger angles of collection of the light. The presence of narrow spectral filters keeps the optimal ratio constant to $1/\sqrt{2}$, independently on the emission angle. Moving away from the thin-crystal regime, walk-off effects must be taken into account, if present, and generically the optimal ratio increases to maintain the overlap between the beams in the process: we studied this aspect in Sec.\ \ref{sec:numerical}, where we also included numerical simulations for type-I SPDC, in  a BBO crystal. To perform our numerical simulations, we used the formulas derived  in Sec.\ \ref{sec:par_appr}, which can also be used to simulate more generic SPDC processes.

Our quantitative analysis of the brightness as a function of emission angle and walk-off effects enriches discussions of other previous work, such as \cite{Bennink_note:_2010}, which is suitable for collinear emission and periodically polarized crystals. However, our work is also a natural starting point for many other analyses. A first obvious generalization would be to consider non-degenerate emission or fiber collecting modes with different waists between the signal and idler. Using both analytical and numerical approaches, as in this paper, it should be possible to derive the associated optimal focus conditions. Moreover, it would be of great interest to extend the study to other quantities such as the heralding ratio or the spectral purity, as discussed in \cite{Bennink_note:_2010,PhysRevA.90.043804}. A careful analysis of these physical properties would lead to a much deeper control of a variety of SPDC configurations and would allow the focusing conditions to be adapted to different experimental requirements.

\begin{acknowledgments}
We thank Costantino Agnesi for his contribution during the initial phase of the project. Part of this work was supported by Ministero dell’Istruzione, dell’Università e della Ricerca (MIUR) 
under the initiative “Departments of Excellence” (Law
232/2016) and by Agenzia Spaziale Italiana (2020-19-HH.0 CUP:
F92F20000000005, Italian Quantum CyberSecurity I-QKD).
\end{acknowledgments}

\appendix
\section{Expansion of longitudinal component of the wavevector}\label{sec:long_comp}

In this appendix we report the expansion of the longitudinal component $k_z$ of the wavevector 
around a particular value of the transverse
components, inside an uniaxial crystal. We will obtain our formulas in the approximation of linear optics.
We recall that we have chosen the $y$ axis so that the optical axis lies on the $(y,z)$ plane.
We also define $\theta$ as the angle between the optical axis of the crystal and the $z$ axis; $n_o$ and $n_e$ are the 
ordinary and extraordinary indices of refraction.

Let us call $(x'=x, y',z')$ the frame of reference associated with the principal axes of the crystal.
For an ordinary wave, i.e.,  polarization along the plane $(x,y')$, the relation between the different components of the wavevector is \cite{Saleh}
\begin{equation}\label{eq:ord_wave}
(k_z')^2+(k'_y)^2+k_x^2=n_o^2 \frac{\omega^2}{c^2} \ .
\end{equation}
For an extraordinary wave instead the relation between $k_z$ and the transverse components is found by solving the equation \cite{Saleh}
\begin{equation}\label{eq:ext_wave}
\frac{(k'_z)^2}{n_o^2}+\frac{(k'_y)^2+k_x^2}{n_e^2}=\frac{\omega^2}{c^2} \ .
\end{equation}
In our computation, we want to rewrite everything in terms of the basis $(x,y,z)$ associated with the pump beam. This can be conveniently done by using the relations \begin{align}
    k'_z&=\cos\theta k_z+\sin\theta k_y \\ 
k'_y&=-\sin\theta k_z+\cos\theta k_y
\end{align}
which, plugged into \eqref{eq:ord_wave},
give
\begin{equation}
\label{eq:kz_ord}
    k_z=\sqrt{\left(n_o^2\frac{\omega}{c}-k_x^2-k_y^2 \right)}
\end{equation}
for the ordinary wave and plugged into \eqref{eq:ext_wave} give
\begin{equation}
\label{eq:kz_ext}
\begin{split}
k_{z}(\kk,\omega)
&=n_\theta^2\left(\frac{1}{n_e^2}-\frac{1}{n_o^2} \right) k_y \sin \theta \cos \theta
+ \\ 
&+\sqrt{\left(\frac{n_\theta\omega}{c}\right)^2-\left(\frac{n_\theta}{n_e} k_x \right)^2-\left(\frac{n_\theta^2}{n_e n_o}k_{y}\right)^2}
\end{split}
\end{equation}
for the extraordinary wave, where we defined
\begin{equation}
    \frac{1}{n_\theta^2}=\frac{\sin^2 \theta}{n_e^2}+\frac{\cos^2 \theta}{n_o^2} \ .
\end{equation}
Equations \eqref{eq:kz_ord} and \eqref{eq:kz_ext} can be encoded in a single expression as in \eqref{kz_general}.

Let us suppose now that
the wave function is peaked around the
transverse
components $\bar{\kk}_{a}=(\overline{k}_{ax},\overline{k}_{ay})$, with $a=\p,\ii,\s$.
In the paraxial approximation discussed in Sec.\ \ref{sec:par_appr}, we want to expand $k_{za}$  in powers of
 $\delta \mathbf{k}_a=\kk_a-\overline{\kk}_a$ up to 
 the second order, namely,
\begin{equation}
\label{kz_expansion}
k_{az} 
=\overline{k}_{az}+\delta \mathbf{k}_a^\top {\bf K}_{1a}
+\frac12\delta \mathbf{k}_a^\top {\bf K}_{2a}\delta \mathbf{k}_a+\mathcal O(\delta \mathbf{k}_a^3) \ .
\end{equation}
Here $\overline{k}_{az}(\omega)\equiv k_{az}(\overline{\kk},\omega)$, while ${\bf K}_{1a}$ and ${\bf K}_{2a}$ collect the contributions of the first and the second derivatives in the expansions, having components
\begin{equation}
{\rm K}_{1a}^{\mu}=
\left.\frac{\partial k_{az}}{\partial k_{a\mu}}
\right|_{\kk_a=\overline{\kk}_a}\,,
\qquad
{\rm K}_{2a}^{\mu\nu}=
\left.\frac{\partial^2 k_{az}}{\partial k_{a\mu}\partial k_{a\nu}}\right|_{\kk_a=\overline{\kk}_a}
\end{equation}
with $\mu=x,y$. The explicit forms of  ${\bf K}_{1a}$ and ${\bf K}_{2a}$ are given by
\begin{align}\label{eq:k1_k2}
{\KK_{1a}}&=
\frac{1}{\beta \overline{k}_{ay}-\overline{k}_{az}}
\begin{pmatrix}
\gamma^2 \overline{k}_{ax}
\\
\left(\beta^2+\gamma^2 \frac{n^2}{n_o^2}\right) \overline{k}_{ay}
-\beta \overline{k}_{az}
\end{pmatrix} \ , 
\\
\KK_{2a}&=
\frac{(\gamma n/n_o )^2}{(\beta \overline{k}_{ay}-\overline{k}_{az})^3}
\begin{pmatrix}
(\frac{n_o\omega}{c})^2-\gamma^2\overline{k}_{ay}^2
& \gamma^2\overline{k}_{ax}\overline{k}_{ay}
\\
\gamma^2\overline{k}_{ax}\overline{k}_{ay}
& (\frac{n\omega }{c})^2
-\gamma^2\overline{k}_{ax}^2
\end{pmatrix}
\end{align}
with $\gamma$, $\beta$ and $n$ defined as in Eq.\ \eqref{kz_general}.

\section{Gaussian beam in dielectric media}\label{sec:gaussian}
In this appendix we consider the Gaussian beam \eqref{eq:spat_wave} propagating in a dielectric medium in order to clarify the physical meaning of certain quantities introduced in the main text. We call $z$ the direction of propagation and consider the beam factorized in the $x$ and $y$ components so that, without loss of generality, we can focus on a generic $\mu$ component
of the field. In the momentum space we have a distribution
\begin{equation}
u(k_{\mu})=
\left( \frac{{\sf w}_{\mu}}{\sqrt{2\pi}}\right)^{\frac{1}{2}}
e^{-\frac{1}{4}{\sf w}_{\mu}^2 (k_{\mu}-k_{0\mu})^2} \ .
\end{equation}
The corresponding field in real space is given by
\begin{equation}\label{realspace}
u(\mu,z)  = \left(\frac{{\sf w}_{\mu}}{(2\pi)^{3/2}}\right)^{\frac{1}{2}}
 \int  {\rm d} k_{\mu} \, e^{-\frac{1}{4}{\sf w}_{\mu}^2(k_{\mu}-k_{0\mu})^2
-i(\mu k_{\mu}+z k_{z})} .
\end{equation}
 This integral expression is rather complicated due to the non trivial dependence of $k_{z}$  on $k_{\mu}$, as shown in \eqref{kz_general}. However, we can use a paraxial approximation, similar to that described in Sec.\  \ref{sec:par_appr}. The point is that the integrand in \eqref{realspace} is peaked around $k_{0\mu}$, so we can expand the longitudinal component $k_{z}$ as (with $\delta k_{\mu}=k_{\mu}-k_{0\mu}$):
\begin{equation*} 
\begin{split}
&u(\mu,z)= \frac{1}{\sqrt{2 \pi}}
\left( \frac{{\sf w}_{\mu}}{\sqrt{2\pi}} \right)^{1/2} \times \\
& \times \int {\rm d}k_\mu \,
e^{-\frac{1}{4}{\sf w}_{\mu}^2(k_\mu-k_{0\mu})^2}  e^{-i\mu k_\mu}
 e^{-i (k_{0z}+\delta k_\mu{\rm K}^\mu_1
+\frac12\delta k_\mu^2{\rm K}^{\mu\mu}_2)z}
\end{split}
\end{equation*}
where we introduced the first and second derivatives of $k_z$, ${\rm K}^{\mu}_1$ and ${\rm K}^{\mu\mu}_2$ respectively, evaluated on $k_{0\mu}$. The resulting Gaussian integral can be performed, obtaining
\begin{equation}\label{realspace_paraxial}
\begin{split}
u(\mu,z,\omega)= & \left(\frac{2}{\pi}\right)^{\frac{1}{4}}\sqrt{\frac{{\sf w}_\mu}{({\sf w}_{\mu}^2+2 i {\rm K}^{\mu \mu}_2 z)}} \ \times \\
&\times \exp \left(-i k_{0\mu}\mu-i k_{0z}z-\frac{(\mu+{\rm K}^\mu_1 z)^2}{{\sf w}_{\mu}^2+2 i {\rm K}^{\mu \mu}_2 z} \right) \ .
\end{split}
\end{equation}
In this way we have found a Gaussian beam with the direction of propagation described by the two wave vectors $k_{0\mu}$ and $k_{0z}$. We stress that this interpretation holds for small values of the $\mu$ component of the wave vector, when the Gaussian profile of the absolute value of \eqref{realspace_paraxial} can be considered perpendicular to the propagation direction. The beam is characterized by the Siegman $q$-parameter in the dielectric media\footnote{The $q$-parameter defined here differs from the standard $q$-parameter defined
in \cite{Saleh} by a factor $-i {\sf w}_\mu^2/z_{0\mu}$, with $z_{0\mu}$ the Rayleigh range.}
\begin{equation}
\label{qm_appendix}
q_{\mu}=
{\sf w}_{\mu}^2+2i {\rm K}_{2}^{\mu \mu}z
\end{equation}
and the Rayleigh range $z_{0\mu}$ is given by
\begin{equation}\label{eq:rayleigh}
z_{0\mu}=-\frac{{\sf w}_\mu^2}{2{\rm K}_2^{\mu \mu}}\, .
\end{equation}
Note also that when $k_{0\mu}=0$, Eq.\ \eqref{realspace} differs from a Gaussian beam propagating in the vacuum in the $z$ direction for the shifting term $-{\rm K}_1^{\mu} z$ in the exponential. In a birefringent medium, this term could be non-zero for an extraordinary wave. 

It is convenient, for a generic field, to introduce the  focal parameter $\xi_{a\mu}$ and
the deviation parameter $\nu_{a\mu}$ as in \eqref{eq:parameters}:
\begin{equation}\label{eq:nu_app}
\xi_{a\mu}(\omega)=-\frac{L}{{\sf w}^2_{a\mu}}{\rm K}_{2a}^{\mu \mu}
\,,\quad
\nu_{a\mu}(\omega)=-\frac{L}{2{\sf w}_{a\mu}}{\rm K}_{1a}^{\mu} \ .
\end{equation}
The parameter $\xi_{ax}$
can also be thought of as the ratio between the length $L$ of the crystal and the Rayleigh range \eqref{eq:rayleigh}; hence it gives information about the focusing condition of the beam in the $\mu$ direction, compared with $L$. 
The quantity $\nu_{a\mu}$, instead, parameterizes the transverse walk-off of the beam: As is clear from \eqref{realspace_paraxial}, at the output of the crystal, the beam is shifted by ${\sf w}_{a\mu}\nu_{a\mu}$ in the $\mu$ direction.
Finally note that, with these definitions, we can write the $q$-parameter as
\begin{align}\label{eq:q_m}
q_{a\mu}(Z)&=
{\sf w}_{a\mu}^2\left(1-i Z \xi_{a \mu}\right)\ ,
\end{align}
where we also substituted $Z=2z/L$.

\section{bi-photon wavefunction in paraxial approximation} \label{sec:computations}

The expression \eqref{kz_expansion} gives the expansion of $k_{az}(k_{ax},k_{ay})$ around the values $\bar{k}_{ax}$ and $\bar{k}_{ay}$. To obtain an approximation of the wavefunction \eqref{PhiGeneral} we should also consider the expansion, around $\overline{\kk}_{\ii}$ and $\overline{\kk}_{\s}$, of the phase mismatch $\Delta k_z$.  Up to the second order in $\boldsymbol\kappa_{\mu}=
(\delta {k}_{\ii \mu},
\delta {k}_{\s \mu})$, we can write
\begin{equation}
\Delta k_{z}
 \simeq
 \Delta \overline{k}_{z}+\sum_{\mu}{\boldsymbol\kappa}_\mu \cdot  {\bf D}_1^\mu+\frac12 \sum_{\mu \nu}{\boldsymbol\kappa}_\mu^\top \cdot {\bf D}_2^{\mu\nu} \cdot {\boldsymbol\kappa}_\nu
\end{equation}
where 
\begin{align*}
&\Delta \overline{k}_{z}=
\Delta k_{z}(\overline{\kk}_{\s},\omega_\s;\overline{\kk}_{\ii},\omega_\ii) \ ,
\\
&{\bf D}_{1}^\mu =
\begin{pmatrix}
{\rm K}_{1\p}^{\mu}-{\rm K}_{1\ii}^{\mu}
\\ 
{\rm K}_{1\p}^{\mu}-{\rm K}_{1\s}^{\mu}
\end{pmatrix}
, \,
{\bf D}_2^{\mu \nu}=
\begin{pmatrix}
{\rm K}_{2\p}^{\mu\nu}-{\rm K}_{2\ii}^{\mu \nu} & {\rm K}_{2\p}^{\mu \nu} \\
{\rm K}_{2\p}^{\mu \nu} & {\rm K}_{2\p}^{\mu \nu}-{\rm K}_{2\s}^{\mu \nu}  \\
\end{pmatrix}  \ .
\end{align*}
In deriving the previous expression we noted that, using $k_{\p\mu}=k_{\ii \mu}+k_{\s \mu}$, 
\begin{equation}
    {\rm K}_{1\p}^{\mu}\equiv \frac{\partial k_{\p z}}{\partial k_{\p \mu}}=\frac{\partial k_{\p z}}{\partial k_{\ii \mu}}=\frac{\partial k_{\p z}}{\partial k_{\s \mu}}
\end{equation}
and similar relations hold for ${\rm K}_{2 \p}^{\mu\nu}$.
Then, to apply the paraxial approximation discussed in Sec.\ \ref{sec:par_appr}, we recast the spatial mode overlap contribution
 \beq 
 \mathcal S=u_\p(\kk_\ii+\kk_\s) u_{\s}^{*}(\kk_{\s},\omega_\s)u_{\ii}^{*}(\kk_{\ii},\omega_\ii)
 \eeq
as
\begin{equation}
\mathcal S=\prod_{a} \left( \frac{{\sf w}_{ax} {\sf w}_{ay}}{2\pi}\right)^{1/2} \prod_{\mu=x,y}e^{-\frac12{\bf C}^\mu_0-
\boldsymbol\kappa^{\top}_\mu \cdot 
{\bf C}_1^\mu
-\frac12\boldsymbol\kappa^{\top}_\mu \cdot {\bf C}_2^{\mu\nu}\cdot \boldsymbol\kappa_\nu}
\end{equation}
with
\begin{align}
& {\bf C}_{0}^{\mu}=
\frac12 \overline{{\sf w}}_{\mu}^2(k_{0 \ii \mu}+k_{0 \s \mu})^2 \ , \qquad
{\bf C}_{1}^{\mu}=
0 \ , \\
& {\bf C}_2^{\mu\mu}=
\frac12
\begin{pmatrix}
{\sf w}_{\p\mu}^2  +
{\sf w}^2_{\ii\mu}
&
{\sf w}^2_{\p \mu}
\\
{\sf w}^2_{\p \mu}
&
{\sf w}^2_{\p \mu} +
{\sf w}^2_{\s \mu}
\end{pmatrix} 
\end{align}
and ${\bf C}^{\mu \nu}=0$ when $\mu \neq \nu$. If we now define, for $j=1,2$,
\beq\label{eq:def_M}
{\bf M}_j(z)=
{\bf C}_j+iz{\bf D}_j
\eeq
we obtain 
\begin{align}\label{eq:phi_gen}
& \Phi(\omega_\ii,\omega_\s)
=
\frac{L}{2}(2\pi)^2 \left( \prod_{a}
\sqrt{\frac{{\sf w}_{ax}{\sf w}_{ay}}{2\pi}}\right) \times \\
& \notag \times
\int_{-1}^{1}{\rm d} Z\,\,
e^{-\frac{iL\Delta \overline{k}_{z}}{2}Z}
\left[\frac{
e^{-{\bf C}_0^x-{\bf C}_0^y+
{\bf M}_{1}(Z)^\top
{\bf M}_{2}(Z)^{-1}
{\bf M}_{1}(Z)}
}
{\text{det}[{\bf M}_{2}(Z)]}\right]^{\frac12}
\end{align}
after solving the Gaussian integral in the transverse components $\mathbf{k}_{\ii}$ and $\mathbf{k}_\s$.
This result was obtained in \cite{kolenderski_note:_2010}, with a different choice of expansion, as explained in Sec.\ \ref{sec:par_appr}. Introducing for each beam $a=\p,\ii,\s$ and for the $x$ and $y$ directions the parameters $q_{a \mu}$ and $\nu_{a \mu}$ of \eqref{eq:q_m} and \eqref{eq:nu_app}, the explicit forms of the matrices are
\begin{equation}
{\bf M}_1=
\begin{pmatrix}
{\bf M}_1^x \\
{\bf M}_1^y
\end{pmatrix} \ , \qquad {\bf M}_2= \left(
\begin{array}{c|c}
\mathbf{M}_2^{xx} & \mathbf{M}_2^{xy} \\ \hline
\mathbf{M}_2^{xy} & \mathbf{M}_2^{yy}
\end{array} \right)
\end{equation}
where
\begin{equation}
    \mathbf{M}_1^{\mu}=i Z
    \begin{pmatrix}
       {\sf w}_{\ii \mu}\nu_{\ii \mu}-{\sf w}_{\p \mu} \nu_{\p \mu} \\
         {\sf w}_{\s \mu}\nu_{\s \mu}-{\sf w}_{\p \mu} \nu_{\p \mu}
    \end{pmatrix}
\end{equation}
and 
\begin{equation}
   \mathbf{M}_2^{\mu\nu} =i \frac{ZL}{2} \mathbf{D}_2^{\mu \nu} \qquad \text{if} \quad \mu \neq \nu \ , 
\end{equation}
\begin{equation}
\mathbf{M}_2^{\mu\mu} = 
 \frac12
\begin{pmatrix}
        q_{\p \mu}(Z)+q_{\ii \mu}^*(Z) & q_{\p \mu}(Z) \\
        q_{\p \mu}(Z) & q_{\p \mu}(Z)+q_{\s \mu}^*(Z) 
\end{pmatrix} \ . 
\end{equation}

 The previous expressions can be simplified by choosing a particular plane of emissions, namely, the $(x,z)$ or the $(y,z)$ planes. In these cases, the integrand in \eqref{eq:phi_gen} is factorized in the $x$ and $y$ components, since this happens for the matrices $\mathbf{D}_2$ and $\mathbf{M}_2$.
The resulting expression is given in \eqref{eq:phi_fact_text}.

\bibliography{optimalbib}

\end{document}